\documentclass[10pt,letterpaper,journal,compsoc]{IEEEtran}
\usepackage{amsmath}
\usepackage{amsfonts}
\usepackage{amssymb}
\usepackage{comment}
\newtheorem{defn}{Definition}

\usepackage[usenames,dvipsnames]{color}
\usepackage[pdftex]{graphicx}
\usepackage{wrapfig}
\usepackage{subcaption}
\usepackage{colortbl}
\usepackage{multicol}

\usepackage[ruled,vlined]{algorithm2e}
\renewcommand{\CommentSty}[1]{\textnormal{#1}}
\DontPrintSemicolon
\SetKwComment{tcp}{$\triangleright$ }{}
\SetVlineSkip{0cm}
\usepackage{hyperref}
\usepackage{breakurl}
\usepackage{url}
\makeatletter \def\url@leostyle{\@ifundefined{selectfont}{\def\UrlFont{\sf}}{\def\UrlFont{\scriptsize\ttfamily}}} \makeatother
\renewcommand{\algorithmcfname}{Alg.}%

\let\emptyset\varnothing

\newcommand*{\fb}{\textcolor{black}}
\newcommand*{\rev}{\textcolor{black}}

\begin{document}

  \title{Fair Task Allocation in Crowdsourced Delivery}

  \author
  {
    Fuat Bas{\i}k$^{\dag*}$\thanks{* Part of this work was done while the author was an intern at IBM.}, Bu\u{g}ra~Gedik$^\dag$, Hakan~Ferhatosmanoglu$^\S$, Kun-Lung Wu$^\ddag$\\
    $\dag$ \textsf{Department of Computer Engineering, Bilkent University, Turkey}\\
    $\S$ \textsf{Department of Computer Science, University of Warwick, UK}
    $\ddag$ \textsf{IBM Research, New York, USA}\\
    \textsf{fuat.basik@bilkent.edu.tr}, \textsf{bgedik@cs.bilkent.edu.tr}, \textsf{hakan.f@warwick.ac.uk}, \textsf{klwu@us.ibm.com}\\
  }

   \markboth{IEEE TRANSACTIONS ON SERVICES COMPUTING,~VOL.~XX, NO.~XX,~XXX~2018}
           {Bas{\i}k et al.: Fair Task Allocation in Crowdsourced Delivery}

  \IEEEcompsoctitleabstractindextext{\begin{abstract}
Faster and more cost-efficient, crowdsourced delivery is
needed to meet the growing customer demands of many industries,
including online shopping, on-demand local delivery, and on-demand transportation.
The power of crowdsourced delivery stems from the large number of workers potentially 
available to provide services and reduce costs. 
It has been shown in social psychology literature that
\emph{fairness} is key to ensuring high worker participation. 
However, existing assignment solutions fall short on modeling the dynamic fairness metric. 
\fb{In this work, we introduce a new assignment strategy for crowdsourced delivery tasks. 
This strategy takes fairness towards workers into consideration, while maximizing the task allocation ratio.}
Since redundant assignments are not possible in delivery tasks, we first introduce a 2-phase \fb{allocation} 
model that increases the reliability of a worker to complete a given task. To realize the effectiveness of our model in practice, 
we present both offline and online versions of our proposed algorithm called \emph{F-Aware}. 
Given a task-to-worker bipartite graph, F-Aware assigns each task to a worker that \fb{minimizes unfairness,}
while allocating tasks to use worker capacities as much as possible.
We present an evaluation of our algorithms with respect to running time, task allocation ratio (TAR), 
as well as \fb{unfairness} and assignment ratio. 
Experiments show that F-Aware runs around $10^7 \times$ faster than the TAR-optimal solution and allocates 
$96.9\%$ of the tasks that can be allocated by it. 
\fb{Moreover, it is shown that, F-Aware is able to provide a much fair distribution of tasks to workers than the best competitor algorithm.}
\end{abstract}
}

 \maketitle
 \IEEEpeerreviewmaketitle
  \section{Introduction}\label{sec:intro}
\noindent 
Spatial crowdsourcing has emerged as a viable solution for delivery logistics, such as on-demand local delivery, online shopping, and on-demand transportation~\cite{ref:rdb-sc}.
As such, it has attracted significant attention from both the academia
and the industry in recent years. For instance, Amazon utilizes the crowd to 
provide same day shipment of packages from warehouses to customers \footnote{http://flex.amazon.com}. 
Postmates\footnote{http://www.postmates.com}, a company offering on
demand food and delivery, is available all around the US.
Enormous growth of the crowdsourced taxi services
Uber\footnote{http://www.uber.com} and Lyft\footnote{http://www.lyft.com} has
attracted significant interest, resulting in numerous research studies being
conducted about them.

Crowdsourced delivery applications have three stakeholders: customers, workers and the matching platform. 
Customers submit tasks of spatial deliveries to the platform. The platform matches the tasks with the workers' availabilities, and allocates 
workers to tasks considering the spatio-temporal requirements. To support faster and cheaper delivery, spatial crowdsourcing platforms require a critical mass of workers. 
The workers should be attracted by a high income
potential which is possible with a large number of customers. 
This situation drives these platforms into a chicken and egg
problem~\cite{ref:physicalInternet}, \fb{in which a powerful network is necessary to 
attract customers and customers are necessary to engage a powerful network.} 

A negative correlation between job satisfaction and 
worker turnover is naturally expected in crowdsourcing environments. According to a 
study with MTurk workers, a common indicator of positive behavior 
of the employer, hence the job satisfaction, is fairness~\cite{ref:MturkStudy}.
\fb{
Fairness can be defined in the context of anti discrimination laws, equity of
opportunity and equality of outcome~\cite{ref:krisna, ref:disparate}. 
In the context of crowdsourcing, the \emph{distributive} fairness
is particularly relevant~\cite{ref:mainPaper, ref:fairDeal}. This definition seeks fairness
based on the proximity between a worker's own input/output ratio and the input/output ratio of a
referent~\cite{ref:adams}. For example, the workers would expect to be assigned a fair number of tasks that is
proportional to their spatio-temporal matching qualities/availabilities for the tasks. Effect of such fairness expectations on the likelihood of participation
is more than that of considerations of self-interest~\cite{ref:fairDeal}. Hence, fairness needs to be considered as an essential concept for sustaining a powerful crowd with significant participation of workers~\cite{ref:boiney}. }

\fb{
In this paper, we study the problem of fair allocation of \emph{delivery tasks} to workers within the context of spatial crowdsourcing. 
The tasks are associated with receive and delivery locations and time constraints. 
The workers inform the platform about their working status using \emph{availabilities}, i.e., the location and time period they are willing to serve. While the primary objective is to maximize the \emph{task allocation ratio (TAR)}, which is the 
ratio of number of allocated tasks to number of all tasks, we aim to achieve this via a fair distribution 
of tasks to workers. Current approaches focus only on the first objective of maximizing the number of tasks under certain constraints from workers~\cite{ref:kazemi}. While this reduces the use of third party services or employing full time couriers~\cite{ref:physicalInternet}, it ignores fairness and worker satisfaction. This can result in lower engagement and migration to other services.~\cite{ref:mainPaper}.
}

There is a number of challenges to achieve an effective and fair crowdsourced delivery.
First, unlike other crowdsourcing applications, a redundant allocation of
tasks is not possible in the crowdsourced delivery. Redundant allocation improves the reliability of 
task completion by increasing the
number of workers the task is assigned to~\cite{ref:rdb-sc}\cite{ref:kazemi}. 
In delivery tasks, however, only one worker can
be allocated to complete the task. Hence, to increase the reliability of the worker
selection in such tasks, we employ a \emph{2-phase \fb{allocation} model}.
In the first phase, the platform selects a set of nominees among available workers
and the task is offered to a subset of them. In the second phase, the platform
selects one worker among those who accepted the offer. 
This avoids the broadcasting of the offer to all nominees in the first phase to 
forestall spamming of the workers. 

The second challenge is to provide \emph{distributional fairness} 
among workers to ensure participation.
Unless workers and their availabilities are identical, assigning each
worker an equal number of tasks is not a fair distribution, 
as it does not take the user input into consideration. To address this
issue, we \fb{call} the input/output ratio \emph{local assignment
ratio} and set the output of a worker as the amount of revenue she gets from the
system, while the input is the total reward of the offers she has accepted (not
necessarily allocated). \fb{This view allows us to define a technical measure of the global fairness as the 
coefficient of variation, a statistical measure of relative variability, of 
all \emph{local assignment ratios}. A low coefficient of variation is associated with the fairness of allocation.}

Without considering fairness, the task allocation problem can
be reduced to the minimum cost flow (MCF) problem~\cite{ref:kazemi}. However, 
the MCF-based solutions fall short to capture fairness, since every
assignment needs to update its cost matrix. 
We introduce \emph{F-Aware} as a solution to assign tasks to workers in a bipartite graph, 
by minimizing the unfairness locally and allocating tasks to fill worker capacities.

The third challenge is to handle online allocation. \fb{In applications such as online shopping (e.g., with a 3-hour
delivery guarantee), the platform can allocate multiple tasks in mini-batches with no global knowledge of all tasks and availabilities in advance.}
In contrast, in on-demand transportation services, like Uber and Lyft, customers
want to know whether the vehicle is on the way, almost instantly. \fb{Therefore the
allocation should be done at the very moment of the task arrival.}
F-Aware with the 2-phase model is shown to be applicable for offline, online, and mini-batch allocation
strategies.

In summary, this paper makes the following contributions:
\begin{itemize}
\item \textbf{Model.} We introduce a generic task \fb{allocation model} to cover
a variety of crowdsourced delivery scenarios. The 2-phase
\fb{allocation} model increases the reliability of task completion by
double-checking a worker's willingness to complete the tasks. 
\emph{This model} handles the case where a potential worker may refuse the task
even though she is available.

\item \textbf{Algorithm.} We introduce a fairness-aware solution called
\emph{F-Aware}, which locally minimizes unfairness, while targeting maximum task
allocation. The MCF-based algorithms fall short on
modeling the dynamic fairness metric, and are not feasible for the online
scenarios. 

\item \textbf{Online Allocation.} We enhance our 2-phase model to perform online
task allocation. We show that \emph{F-Aware} is effective also for online
and mini-batch allocation scenarios, as it is for offline allocation.
\end{itemize}

\fb{We provide a comprehensive experimental study using real-world datasets to showcase the effectiveness and efficiency
of our 2-phase model and of the \emph{F-Aware} algorithm in terms of running time, 
task allocation ratio, and fairness it achieves.}

The rest of this paper is organized as follows. Section~\ref{sec:problem}
gives the preliminaries of the problem. Section~\ref{sec:model} explains the
details of our 2-phase task \fb{allocation} model. Section~\ref{sec:online} extends
our approach to online task allocation. Section~\ref{sec:experimental}
presents the experimental evaluation. Section~\ref{sec:related} discusses the
related work. Finally, Section~\ref{sec:conclusion} concludes the paper.
  \section{Problem Definition}\label{sec:problem}
\begin{figure*}[t]
\centering
\includegraphics[width=\linewidth]{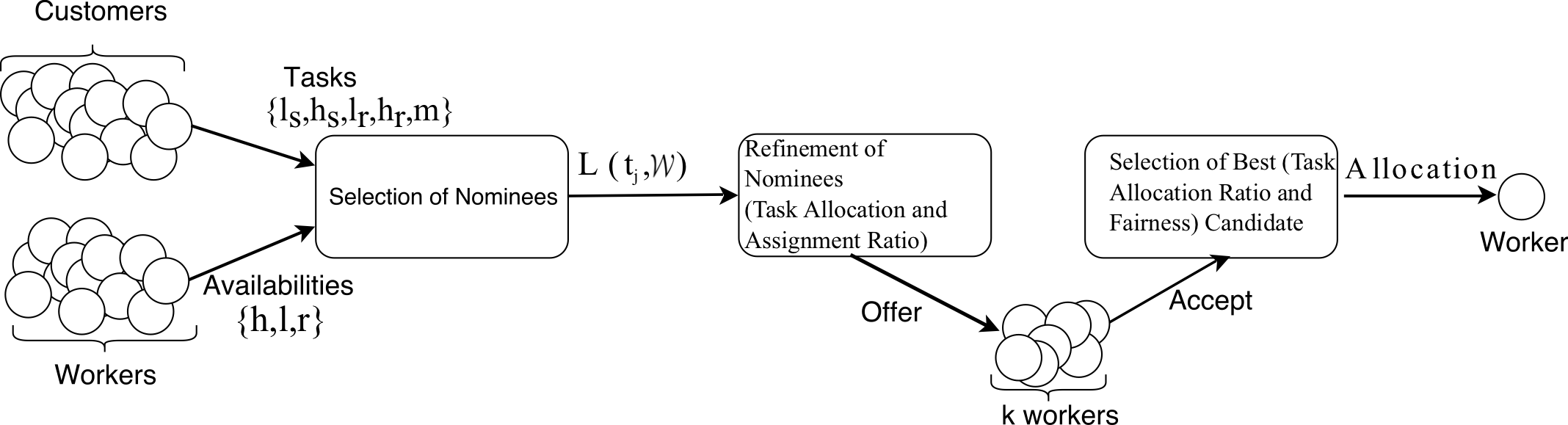}
\caption{Tasks and availabilities are the inputs of the platform. 
For each task,  the nominees are identified. During this calculation, the system also
finds out the \emph{acceptance probability}, based on
Equation~\ref{eq:reliability}. The task is multicasted to $k$ of those
nominees. Workers who accept the offer are referred to as \emph{candidates}
and one worker among the candidates is selected for allocation.}
\label{fig:workflow}
\end{figure*}
\noindent We aim to develop a new strategy on allocation of tasks
to workers in a crowdsourced environment.
The overall goal is to maximize the 
task allocation ratio (TAR), the ratio of number of allocated task over the number of all tasks, while distributing tasks to workers fairly.
We now give the preliminaries of the domain and formalize this multi-criteria optimization problem. 
\begin{defn}{\textbf{Time period.}}
A time period, $h$, is a pair of date-time values $b$ and $e$, representing
the beginning and end times, respectively. 
\end{defn}
\begin{defn}{\textbf{Delivery task.}}
\fb{Tasks are in the form of spatio-temporal deliveries, such as workers need to move to the source of the delivery to receive the item and deliver it to the recipient. In this manner, one can consider a task as a composite of \emph{receive} and \emph{deliver} steps. 
We represent the set of all tasks with $\mathcal{T}$, and $i^{th}$ task with $t _{i}$.
A task is a quintuple $\{h_{s}, l_{s}, h_{r}, l_{r}, m\}.$ Here, $h$ and $l$ represents a time period and a geo-spatial point  such as a latitude/longitude pair, respectively. 
Subscripts $s$ and $r$ stands for the source and the destination of the task.
In other words, $t_{i}.{l_s}$ stands for location of the source while $t_{i}.{l_r}$ stands for location to deliver the item for $i^{th}$ task $t_i$. Note that these steps are associated with different time periods, as \textit{receive} and \textit{deliver} steps have their own validity periods $h_s$ and $h_r$. Lastly, $m$ represents the reward of the task.}
\end{defn}
\begin{defn}\textbf{Worker.} \fb{Workers are people who participate in the
platform to make money. We represent the set of all workers with $\mathcal{W}$, and $i^{th}$ worker with $w _{i}$. Each worker, is a triple $\{A, c, f\}$. 
$w_i.A$ is the set of \emph{availabilities} of worker $w_i$, 
$w_i.c$ is her capacity and $w_i.f$ is the local assignment ratio.} \emph{Local assignment ratio}, which will be detailed later, is a dynamic 
metric used to determine how fair the system treated a worker so far, 
defined as the ratio of the worker's revenue over the total reward of the 
offers she has accepted. \fb{The revenue of the worker equals to sum of rewards of the tasks allocated to her.}
\end{defn}
\begin{defn}\textbf{Availability.} Workers inform the system about their 
working status using \emph{availabilities}. \fb{$a_{ip}$ is the $p^{th}$
\textit{availability} of $i^{th}$ worker $w_i$, such that $a_{ip} \in w_i.A$.
Each availability is a triple, $\{h, l, r\}$.
$a_{ip}.l$ is a geo-spatial point. It is the center of the region in which $w_i$ is willing to accept tasks. $a_{ip}.r$ is the radius of the same region. Let us represent this region with the $Cr(l,r)$ function. 
The worker is ready to serve during the time period $a_{ip}.h$.}
\end{defn}

For a task, $t_i$, to be completed, a worker has to move to the \emph{receive}
location, $t_{i}.{l_s}$, during its validity period $t_{i}.{h_s}$; and after that, she has to move to the \emph{deliver} location, 
$t_{i}.{l_r}$, during its validity period $t_{i}.{h_r}$. 
An example task $t_i$ in crowdsourced delivery could be to 
pick-up a gift item from a local shop ($t_{i}.{l_s}$) between 12:00 and
14:00 on 23rd of December 2016 ($t_{i}.{h_s}$), and deliver it to a home
address ($t_{i}.{l_r}$) between 18:00 and 22:00 on the same day 
($t_{i}.{h_r}$). In return for completing this task, the worker will be paid
$t_i.m$ amount of money. In the real world, multiple parameters effect the
$t_i.m$ value, including the distance between $t_{i}.{l_s}$ and $t_{i}.{l_r}$, 
the size of the package, and the sensitivity of the content, etc.

\subsection{Fairness}
\noindent An effective network is key to building a powerful crowdsourcing platform, therefore, 
providing continuous participation of workers and avoiding worker turnover are crucial. 
A negative correlation is naturally expected between job satisfaction and
worker turnover in crowdsourcing environments. 
According to a study with MTurk~\footnote{http://www.mturk.com/} workers, $11$ to $26$ percent 
of turnover in crowdsourcing environment  is explained with job satisfaction.
In the same study, fairness is listed as one of the most common indicator of the job satisfaction~\cite{ref:MturkStudy, ref:shappe}.
Fairness needs to be considered as a first class citizen in designing
crowdsourcing applications to ensure long term commitment and
participation~\cite{ref:boiney}.

There are three major forms of fairness defined in the social psychology literature, namely: 
\textit{procedural}, \textit{interactional}, and \textit{distributive}.
\textit{Procedural} fairness is the perception of justice 
on the procedures, policies, and the criteria used by the decision
maker~\cite{ref:mainPaper}. \textit{Interactional} fairness is the
interpersonal aspect of the \emph{procedural} fairness. 
\emph{Distributive} fairness is defined as the
proximity between a worker's own input/output ratio and the input/output ratio
of a referent~\cite{ref:adams}. Prior research on the relationship between 
\emph{fairness} and job satisfaction shows that when fairness is regressed along all three
dimensions, the job satisfaction gets impacted the most due to the loss of~\emph{distributed fairness}~\cite{ref:shappe}. 
Note that, unless the workers are identical, 
assigning each worker to an equal number of tasks is not a fair distribution by this definition, 
as it does not take user input into consideration. 
Therefore, we define a new \emph{fairness} model that captures \emph{distributive fairness}, 
which will be detailed shortly.

\subsection{Formalization}
\noindent With the given definitions, let us first define the 
problem before discussing each component separately.\\
\noindent\textbf{Fair allocation of delivery tasks in a crowdsourcing environment:}
Given the set of delivery tasks $\mathcal{T}$ and the set of
workers $\mathcal{W}$, represented with their availabilities, the 
problem is allocating tasks to workers with the goals of maximizing the
task allocation ratio ($TAR$) and minimizing the unfairness 
($\mathcal{F}$) objectives, \rev{under the \emph{candidacy}, \emph{capacity} and 
\emph{assignment ratio (AR)} constraints which will be explained next.}
\medskip\\

\noindent\textbf{Task Allocation Ratio (TAR):} To reduce the dependency of the businesses to using a third party service
we set maximizing \emph{task allocation ratio (TAR)} as the first component of our objective 
function. $TAR$ is defined as the ratio of the
number of allocated tasks over the number of all tasks. Formally, let
$T_{all}$ be the set of allocated tasks and $\mathcal{T}$ be the set of all tasks. 
The $TAR$, defined as:

\begin{equation}
TAR = \frac{|T_{all}|}{|\mathcal{T}|}
\end{equation}

Unlike other crowdsourcing applications, redundant allocation of tasks is not
possible in crowdsourced delivery. Therefore, in order to increase the
reliability of a worker completing a given task, the allocation is done via a
2-phase procedure, illustrated in Figure~\ref{fig:workflow}. In the first
phase, the system nominates a set of workers whose \emph{availabilities} are
suitable to complete the task. The task is offered to these \emph{nominees},
and they have an opportunity to accept or reject it. A worker may refuse the
offer, even though she is available. For brevity, we leave the details of
nomination and offering strategy to the next section. Workers who have
accepted the offer are called \emph{candidates}. In the second phase, one
worker among the candidates is selected and is assigned to the task. 
\medskip\\
\noindent\textbf{Fairness:}
Recall that \emph{distributive fairness} is defined as the
proximity between a worker's own input/output ratio and the input/output ratio
of a referent. In the 2-phase allocation model, the \emph{input} of a worker is 
the \fb{total reward} of the offers she accepted. 
Note that \emph{availabilities} cannot be used as
input since a worker might reject offers even though she is available.
The output of a worker, on
the other hand, is the amount of money she earned. Intuitively, not all tasks
have the same complexity, yet the reward of each task is proportional to its
hardness. To capture the hardness of the tasks while determining a worker's
input/output ratio, instead of counting the number of tasks a worker accepted
or got assigned, we use the reward of each task. Each worker $w_i$ is associated
with a \emph{local assignment ratio (LAR)}, $w.f_i$, defined as 
the ratio of the total reward of tasks allocated to a worker \fb{(output)} over 
the total reward of tasks she has accepted \fb{(input)}. 
Formally, let $w_i.{T_{all}}$ be the set of tasks allocated to 
worker $w_i$, and $w_i.{T_{acc}}$ is the set of offers she has accepted, \fb{the LAR of $w_i$ is defined as:}

\begin{equation}
w_i.f = \frac{\sum_{j=1}^{x} t_j.m}{\sum_{k=1}^{y} t_k.m}, 
\: \forall_{j,k} \: t_j \in w_i.{T_{all}}, \: t_k \in w_i.{T_{acc}}
\end{equation}

where
\begin{equation}
x = |w_i.{T_{all}}| \; and \; y = min(|w_i.{T_{acc}}|,w_i.c)
\end{equation}

The number of tasks a worker accepts can exceed her capacity, since she will
be assigned to a subset of the tasks she has accepted. However, since the
capacity limits the number of allocations, we consider minimum of
$\{|w_i.{T_{acc}}|, w_i.c\}$ acceptances when calculating the denominator. The system
is considered more fair as the proximity of the \emph{LAR} values will be higher.
Although the standard deviation of the set of \emph{LAR} values represents this
proximity, using it as the evaluation metric would be misleading since the 
different allocation schemes will have different number of tasks allocated. Hence,
we evaluate the overall fairness of the system, $\mathcal{F}$, using
coefficient of variation of the set of \emph{LAR} values, i.e., standard
deviation of the LAR values divided by their mean. Let $F$ be the set of all
\emph{local assignment ratio} values of users system fairness, $\mathcal{F}$, is 
formalized as:

\begin{equation}
\mathcal{F} = \frac{\sigma(F)}{\mu(F)}
\end{equation}

\begin{table}
\centering
\caption{Commonly Used Notation}
\label{table:notation}
\begin{tabular}{|p{2.3cm}|p{5.5cm}|}
\hline
\textbf{Notation}  & \textbf{Explanation}                                     \\
\hline
h $[b,e]$    & time period; beginning from $b$, ending at $e$  \\             
\hline
$w \{A, c, f\}$    & worker; consists of set of availabilities ($w.A$), capacity ($w.c$) and local assignment ratio ($w.f$)\\
\hline
$t \{h_s,l_s, h_r, l_r, m\}$    & task; consists of source location ($t.{l_s}$), source time period ($t.{h_s}$), destination location ($t.{l_r}$), destination time period ($t.{h_r}$) and a reward ($t.m$).  \\             
\hline
$a_{ip} \{h,l,r\} $    & $p^{th}$ availability of $i^{th}$ worker $w_i$ consists of a time period $h$, a location $l$ and radius $r$  \\
\hline
$\mathcal{T}$, $\mathcal{W}$, $\mathcal{A}$, $F$ & set of all tasks, workers, availabilities and local assignment ratios respectively.\\
\hline
\begin{tabular}{@{}c@{}}$T_{all}, w_i.{T_{all}},$ \\ $w_i.{T_{acc}}$\end{tabular} & set of allocated tasks, allocated tasks to worker $w_i$ and tasks accepted by $w_i$. \\
\hline
\end{tabular}
\end{table}
\noindent\textbf{Candidacy Constraint:} Recall $w_i.{T_{acc}}$ is the set of tasks worker $w_i$ has accepted.
A given task $t_j$ can be assigned to worker $w_i$ only if the system \textit{nominated} her for the
task, and she \textit{accepted} the offer: $t_i \in w_i.{T_{acc}}$.
\medskip\\
\noindent\textbf{Capacity Constraint:} Definition of the capacity constraint is intuitive. 
The number of tasks assigned to a worker cannot exceed her capacity: $|w_i.{T_{all}}| \leq w_i.c$.
\medskip\\
\fb{
\noindent\textbf{Assignment Ratio Constraint:}
Note that each task is offered to a set of \emph{nominees}, yet among 
the \emph{candidates} who accept the offer, only one worker will be selected. 
To forestall spamming of workers and avoid unnecessary communication costs, 
we avoid broadcasting offers to \emph{nominees}. 
At the one extreme, the task can be repeatedly unicasted
until one \emph{nominee} accepts it. However, this approach would cause
potentially long wait times. Therefore, we present a hybrid solution:
multicasting the offer to $k$ workers, which avoids spamming of the workers
while increasing the probability of at least one nominee accepting the offer.
The value of $k$ is calculated for each task independently. In the next
section, we show that higher values of $k$ will result in higher number of
\emph{candidates}, but it leads to a large set of spammed workers. To be able
to define an upper limit, we introduce a system-wide metric called
\emph{assignment ratio (AR)}, which is the ratio of the \emph{number of
\fb{allocated tasks}} over the \emph{number of \fb{accepted offers}}. 
Higher \emph{AR} indicates more accurate \emph{nominee} selection, or 
less number of spammed workers. Therefore, we constraint assignment ratio to be higher than a 
predefined threshold $\theta$.}
We formalize the \emph{AR} as follows:
\begin{equation}
AR = \frac{|T_{all}|}{\sum_{i=1}^{|\mathcal{W}|} |w_i.{T_{acc}}|}
\end{equation}
\fb{
And define \emph{assignment ratio} constraint as $AR \geq \theta$. Later, we will discuss 
relaxing this constraint to decrease the wait time of the customers.}

  \section{Allocation Model}\label{sec:model}
\noindent In this section, we describe the details of the 2-phase \fb{allocation} model. 
Inputs to the system are the tasks from the customers, and the
\emph{availabilities} from the workers, both with time and location components. 
In many practical cases, both the tasks and the
availabilities are registered in advance. Hence, one needs to check if a worker is still
willing to do the job. Our model is able to do this check
to increase the reliability of the worker with respect to the completion
of the given task. In the next section, we also cover the online \fb{allocation} scenario, in which tasks and
availabilities can appear anywhere, anytime.

In the remaining parts of this section, we give the details of the 2-phase
model in three steps: ($i$) nomination of workers, ($ii$)
batched progressive offer strategy, and ($iii$) task \fb{allocation}.

\subsection{Nomination of Workers}
\begin{figure}[!t]
\centering
\includegraphics[width=.85\linewidth]{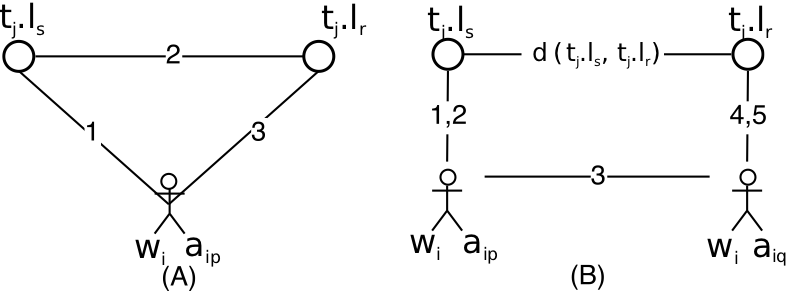}
\caption{Movement cost calculation of a task is different depending 
on the number of availabilities satisfying it. 
Roads are numbered with respect to travel order.}
\label{fig:movement}
\end{figure}

\noindent Initial step of the 2-phase \fb{allocation} procedure is to find out
the appropriate set of workers to offer the task. 
These workers are called \emph{nominees}.
For a worker $w_i$ to be nominated to a task $t_j$, her availability set, 
$w_i.A$, should contain the necessary availabilities to satisfy both steps of $t_j$. 
\fb{Recall since tasks are in the form of spatio-temporal
deliveries, one can divide them into two: \emph{receive} and \emph{deliver} steps.}
Satisfying one of these parts, \fb{lets assume \emph{receive}}, means that worker
$w_i$ should have at least one availability $a_{ip}$, such that time period of task $t_{j}.{h_s}$
intersects with time period of availability $a_{ip}.h$ and source location 
$t_{j}.{l_s}$ lies inside the region $Cr(a_{ip}.l,a_{ip}.r)$.

One can formalize the \fb{satisfaction} relation between a step, \fb{\emph{receive} or 
\emph{deliver}}, of a task $t_j$ and an \emph{availability} $a_{ip}$ of a worker as a function $S$:

\begin{equation}
	\begin{aligned}
	S(t_j.{h_s}, t_j.{l_s}, a_{ip}) \equiv  (t_j.{h_s} \cap a_{ip}.h \neq \emptyset) \wedge \\ 
	(t_j.{l_s} \in Cr(a_{ip}.l,a_{ip}.r))
	\end{aligned}
\end{equation}

The intuition behind dividing the tasks into two steps is that, a worker can
have separate \emph{availabilities} such that one satisfies the requirements
of the \emph{receive} step and the other satisfies the \emph{deliver} step.
Given the set of workers, the system searches for those workers that have
satisfying availabilities for both steps of a task. To formulate this
relation, we employ the $S$ function to locate the pair of availabilities of a
worker $w_i$ that satisfy the \emph{receive} and \emph{deliver} steps,
respectively. We denote the resulting function as $N(t_j, w_i)$. If no such pair
can be located, then the function produces an empty set.

\begin{equation}
N(t_j, w_i) =  \begin{cases}  
      \{a_{ip}, a_{iq}\} & \begin{split} & \exists \, a_{ip} \in w_i.A | S(t_j.{h_s}, t_j.{l_s}, a_{ip})\, \wedge \\
                        & \exists \, a_{iq} \in w_i.A | S(t_j.{h_r}, t_j.{l_r}, a_{iq})  \end{split}\\
      \emptyset & otherwise
   \end{cases}
   \label{eq:nomination}
\end{equation}

\noindent We should note that, in Equation~\ref{eq:nomination}, $p$ and $q$
values can be equal, which means a single \emph{availability} might satisfy
both requirements.

It would be unrealistic to assume that
all nominated workers will have the same probability to accept the offered task.
Besides availability, there are many factors that influence such a decision.
Existing research showed that workers are willing to accept the tasks that are
less costly for them and closer to their home locations~\cite{ref:realWorld}.
Therefore, the \emph{acceptance probability} of each nominated worker for a given task 
is negatively correlated with the task's cost, which is the movement cost in our case.

Many existing local delivery systems use the distance between the source and
the destination as an indicator of the payment amount. It means that, tasks
which require longer traveling pay more to their workers. On the other hand,
the movement cost of a task might differ between workers, as they should move
towards the source from their current location or move back to their previous
location from the destination. Since we do not track the workers' locations,
we assume that a worker is at the location provided as part of her valid
\emph{availability} and she tends to go back to that
location, after completing the \emph{deliver} step.
We also assume that the acceptance does not depend on the previous acceptances or rejections.

Figure~\ref{fig:movement} shows two different scenarios regarding the
calculation of the movement cost of a task. 
Let us define the movement cost of a task as $\alpha +
\beta$, and let $d$ be the function that calculates the distance between two
geo-spatial points. The distance between source and the destination, given by
$\alpha = d(t_{j}.{l_s}, t_j.{l_r})$, remains the same no matter which worker is
assigned to the task. On the other hand, the distance traveled towards the
source and from the destination, $\beta$, depends on the worker
availabilities. In the first scenario, single availability of the
worker, $a_{ip}$, satisfies both of the steps of the task. Therefore,
the worker will only move towards the source, and from the destination 
($\beta=d(a_{ip}.l,t_j.{l_s}) + d(t_j.{l_r},a_{ip}.l)$). In the second scenario
however, after moving towards the source, as well as after delivering the item, 
worker will go back her initial position. Moreover, 
distances between the locations of those \emph{availabilities} should be
considered as well. Consequently, workers movement in the second scenario would be equal to 
$ \beta = 2 \times d(a_{ip}.l,t_j.{l_s}) + 2 \times d(t_j.{l_r},a_{iq}.l)  + d(a_{ip}.l,a_{iq}.l)$.

\fb{The reward of the task} is positively correlated with
$\alpha$ \fb{and indifferent to who completes the task.}
Minimum value for the total travel distance is $2\times\alpha$,
which happens when the worker is already at the source location, or at the
destination location, or lies on the linear line connecting these two points,
i.e. $\alpha = \beta$. Intuitively, workers are willing to accept tasks with
high income and less movement. Therefore, acceptance probability of a task $t_j$
by worker $w_i$, denoted as $R(t_j,w_i)$, is positively correlated with $\alpha$,
but negatively correlated with $\beta$. For the ideal case, $\alpha = \beta$
acceptance probability should be $1$. We formalize this model in
Equation~\ref{eq:reliability}. To take into account the probability of a
worker refusing an offer even when she is a perfect match, we use a constant
$c$, where $0 \leq c \leq 1$.

\begin{equation}
R(t_j,w_i) = e^{\alpha-\beta} \times c
\label{eq:reliability}
\end{equation}

Since $\alpha \leq \beta$ and $0 \leq c \leq 1$ the value of $R$ is guaranteed to
lay between $0$ and $1$. Selection of nominees for a task $t_j$ outputs the set
$L(t_j, \mathcal{W})$, which contains the worker $w_i$ and acceptance probability $R(t_j,w_i)$
pair. This set is used in our batched progressive offer strategy, a technique
we implement to minimize the waiting time, while maximizing the
\emph{assignment ratio}.

\begin{equation}
L(t_j, \mathcal{W}) = \{ w_i, R(t_j,w_i) | w_i \in \mathcal{W} \wedge N(t_j, w_i) \neq \emptyset \}
\end{equation}

\fb{Let $\mathcal{A}$ be the set of all availabilities.}
Brute force approach to construct $L(t_j, \mathcal{W})$ for a given task $t_j \in \mathcal{T}$
iterates through the availabilities set $\mathcal{A}$ and calculates $R(t_j,w_i)$
if $N(t_j,w_i)\neq \emptyset$, for each worker $w_i \in \mathcal{W}$. For all tasks,
complexity of this calculation is $\mathcal{O(|\mathcal{T}||\mathcal{A}|)}$.
To decrease the computation time, we index availabilities on the temporal
dimension using interval trees. For each task, we query this index two times,
one for the \emph{receive} step, and one for the \emph{deliver} step. We check
the spatial intersection only for the resulting availabilities. In the
\emph{Online Task Allocation} section, we discuss how long to wait between task
arrival and nominee selection processes.

\subsection{Batched Progressive Offer Strategy}
\noindent There are three different approaches regarding offering a task to a
predefined set of \emph{nominees}. The first approach is broadcasting the
offer. To forestall spamming of workers and avoid unnecessary communication costs, 
we avoid broadcasting offers to nominees. Moreover, 
we want to minimize the number of cases where a worker accepts the 
incoming offer but is not allocated to the task, as this might cause churn over time. 
Therefore, the ideal case is when there is exactly one
candidate. Reaching this ideal case is only possible with unicasting the offer.
However, there is the probability of the nominee rejecting the offer, even
though she is available. Therefore, the system should follow a progressive
approach, by offering the task one by one, until somebody accepts, and waiting
for a preset time between each round. Obviously, this will result in a
long waiting time for the customer to see if his task is going to be served or
not. To avoid both situations, we multicast the offer to $k$ nominees in
batches, progressively, until there is at least one candidate. With this
approach, we decrease the waiting time of the customer, while limiting the
number of candidates. For each task, once the value of $k$ is calculated, it
is used for further batching, if necessary.

Let us call the probability of the task being accepted by at least one nominee as
\emph{probability of response}. For each task $t_j$, the probability of response,
when offered to $k$ nominees, can be calculated using the probability of
all $k$ nominees refusing it. Recall that $L(t_j, \mathcal{W})$ is the set of nominee and
\emph{acceptance probability} pairs, and assume that it is sorted by the
\emph{acceptance probability} values. Probability of response for a 
batch of $k$ workers is calculated as follows.

\begin{equation}
P(k, L(t_j, \mathcal{W})) = 1 - \prod_{i=1}^{k} (1- (R(t_j,w_i)))
\label{eq:acceptance}
\end{equation}
By keeping the value of $P(k, L(t_j, \mathcal{W}))$ above a tuning parameter $\epsilon$, the
lower bound of the $k$ can be defined. We call $\epsilon$, the \emph{threshold
of probability of response}.

The upper bound, however, is calculated using the \emph{assignment ratio} as a
constraint. To limit the value of $k$, first we should be able to predict how
many nominees are likely to accept the offer. In the worst case for assignment ratio all $k$
nominees accept the offer. As we try to maximize the \emph{assignment ratio}, the number
of candidates should not exceed a certain number. Although it is not realistic
to expect that all $k$ workers will accept the offer, it is still useful to
limit the $k$ value.

Given a task $t_j$, let us define the probability of $i^{th}$ worker accepting
the offer as a random variable $x_i$. Then the expected value of it is
$E[x_i]= R(t_j,w_i)$. Transforming into multiple workers, assignment ratio would
be one over the number of candidates, therefore, it would be equal to
$E[\frac{1}{x_1 + x_2 + ... + x_k}]$. From probability theory, it is known
that $E[\frac{1}{x_1 + x_2 + ... + x_k}] \geq \frac{1}{E[x_1 + x_2 + ... +
x_k]}$, thus the latter could be used as a lower bound for the expected value,
where, $E[x_1 + x_2 ... + x_k] = R(t_j,w_1) + R(t_j,w_2)+ ... R(t_j,w_k)$. With this
at hand, let us define a function $\mathcal{E}(k,L(t_j))$ as a lower bound for the
expected value of the \emph{assignment ratio}:

\begin{equation}
\mathcal{E}(k, L(t_j, \mathcal{W})) = \frac{1}{E[x_1+x_2+ ... + x_k]} 
\end{equation}

Recall one of the constraints is keeping the \emph{assignment ratio} of the system
above a predefined threshold $\theta$. We use the same threshold, 
as a lower bound to function $\mathcal{E}(k, L(t_j, \mathcal{W}))$ to satisfy assignment 
ratio constraint locally, for each allocation. Although local satisfaction is 
stronger than the global constraint, as we will discuss next, the assignment 
ratio constraint is relaxed when it contradicts with the probability of response.

While the probability of response $P(k, L(t_j, \mathcal{W}))$ is positively correlated with $k$, 
as it is shown above, the expected value of the \emph{assignment ratio} 
(value of $\mathcal{E}(k, L(t_j, \mathcal{W}))$) decreases with it. 
Therefore, bounding $\mathcal{E}$ function from below, sets an upper bound on the value of $k$.
With the above defined thresholds, the value of $k$ should guarantee that probability of response 
$P(k, L(t_j, \mathcal{W}))$ is above $\epsilon$, while the assignment ratio constraint 
is satisfied for each task, i.e. $\mathcal{E}(k, L(t_j, \mathcal{W}))$ is above $\theta$.

In summary, $k$ is selected using following inequality:
\begin{equation}
	\begin{aligned}
		k &\geq min\{k | (P(k, L(t_j, \mathcal{W})) \geq \epsilon\} \\ 
		k &\leq max\{k | \mathcal{E}(k, L(t_j, \mathcal{W})) \geq \theta\}
	\end{aligned}
\label{eq:kequality}
\end{equation}

$k$ is set to the maximum value that satisfies both of those inequalities.
However, the given inequality might be invalid with respect to selection of the
$\epsilon$ and $\theta$ values. Consider that, even the smallest $k$ value
satisfying the upper inequality might not satisfy the lower one. 
In that case, \fb{we relax the assignment ratio constraint}, 
and use the $k$ that satisfies the upper inequality.

After this calculation, the task is offered to the first $k$ workers and the
system waits for a predefined time. Recall that, workers are sorted in the decreasing
order of the acceptance probability. In case of all nominees refuse the offer,
the task is offered to the next $k$ workers, until there is at least one
candidate or all nominees are asked.

\subsection{Task Allocation}
\begin{figure}[!t]
\centering
\includegraphics[width=.65\linewidth]{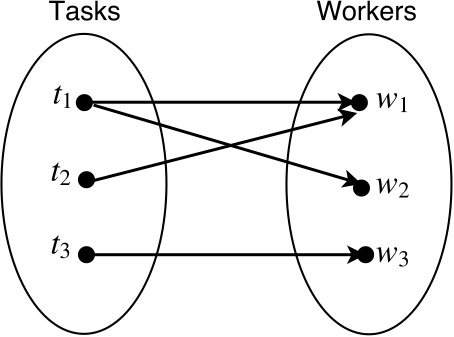}
\caption{Allocation of tasks to candidates}
\label{fig:tcr}
\end{figure}

\noindent The last step of the 2-phase model is to select one worker among
the candidates to allocate the task. 

Figure~\ref{fig:tcr} shows an example scenario in which there are three tasks
and three candidates. The edges are from tasks to their candidates. For
example $t_1$ is accepted by $w_1$ and $w_2$, while $w_1$ is the only
candidate for $t_2$. The simplest version of this problem is finding a
one-to-one assignment scheme of tasks to workers, assuming all the edges have
equal weight. By definition, this is the bipartite graph assignment problem.
In our specific case, there is a one-to-many relationship between the tasks
and the workers. Moreover, to capture the spatial aspect of the problem, one
might want to use edge weights to represent movement cost and add capacity to
workers as well. Including these additional constraints, the task-to-worker
bipartite graph can be represented as a flow graph by adding a source and a
sink. With this representation, task to worker allocation can be reduced to
the minimum cost flow (MCF) problem and solved optimally with well known
algorithms, i.e. successive shortest path or cycle
canceling~\cite{ref:kazemi}.

Even though they reach optimal result on task allocation, minimum cost
flow based solutions have their own drawbacks. First of all, running time of
the optimal solution is far from being feasible because of high complexity.
Given the task to candidate bipartite graph $G(E,V)$, s.t. $V = \mathcal{T}
\cup \mathcal{W}$ and $E$ represents the acceptances, the complexity of
 a careful implementation of successive shortest path algorithm is $\mathcal{O}(|V||E|log|V|)$.
Second, with the aforementioned definition of 
\emph{fairness}, it is hard to integrate it into the MCF-based algorithms. 
This is because at each iteration \emph{LAR} values are updated, which updates
the cost-matrix as well. 

\rev{Assuming that \emph{LAR} values are static, introducing fairness as a new constraint 
transforms the problem to minimum cost maximum flow problem. 
The goal of this problem is to select the minimum cost flow among multiple maximum flows. 
In our setting, this corresponds to maximizing task allocation ratio while minimizing unfairness objectives.
This problem could be solved with Hungarian algorithm, however, similar to the previous discussion, 
the high complexity of the algorithm, $\mathcal{O}(n^3)$, 
makes it inefficient for online allocation scenarios~\cite{ref:hungarian}.
}


The na\"{i}ve approach for allocation is random selection of one candidate,
which is used as one of the baselines in our experimental evaluation. On the
other hand, to capture the spatial aspect of crowdsourced delivery, it is
beneficial to allocate the nearest worker~\cite{ref:kazemi}. This approach can
be extended by proactive allocation of workers, if distribution of tasks is
known in advance~\cite{ref:fairAllocation}. Last but not the least, inspired
from file allocation techniques from the operating systems literature,
selecting the Least Allocated Worker first could increase the \emph{Task
Allocation Ratio} by reserving room for further allocations. While all
approaches are feasible in terms of their running time, our experimental
evaluation shows that they either fell short in terms of modeling the fairness
or have subpar performance with respect to task allocation when the capacity
is constrained.

\begin{algorithm}[!t]
	\begin{small}
 	\KwData{ $G(E,V)$: Task-Worker Bipartite Graph s.t. $ V = \mathcal{T} \cup \mathcal{W}$}
 	\KwResult{$G'(E,V)$:Updated Bipartite Graph}
 	$G'(E,V) \leftarrow G(E,V)$ \tcp*{Copy the original graph}
 	$\mathcal{T} \leftarrow sort(\mathcal{T})$
 	\tcp*[f]{Tasks sorted in increasing order of in-degree}\\
 	\For(\tcp*[f]{For each task in task list}){$t \in \mathcal{T}$}{
 	 	\tcp{Get the subset of workers sharing an edge with $t$}
 		$\mathcal{W'} \leftarrow G(E,V).get(t)$\\
 		$\mathcal{W'} \leftarrow sort(\mathcal{W'})$
 		\tcp*[f]{Workers sorted in increasing LAR order}\\
 		$isAssigned \leftarrow False$\\
 		$index \leftarrow 0$\\
 		\While(\tcp*[f]{Assignment is not done but there are still candidates}){$ \neg isAssigned \,\&\&\, index < |\mathcal{W'}|$}{
	 		$w \leftarrow W'.get(index)$\\
	 		\If(\tcp*[f]{If worker has capacity}){$w.c > 0$}{
	 			$G'(E,V) \leftarrow G'(E,V) \cup assign(t,w)$ \tcp*{Mark edge}
	 			$profit(w) \leftarrow profit(w) + t.m$ \tcp*{Update profit}
	 			$w.c \leftarrow w.c -1$  \tcp*{Decrease capacity}			
	 			$isAssigned \leftarrow True$
	 		}\Else(){
	 			$index \leftarrow index + 1$
	 		}
	 	}
	}
		\textbf{return} $G'(E,V)$ \tcp*{Return the modified bipartite graph}
	 \caption{F-Aware Task Allocation}
  	\label{algo:fAware}	
	\end{small} 	
\end{algorithm}

To cope with the aforementioned challenges, we introduce an algorithm called
\emph{F-Aware}, given in
Algorithm~\ref{algo:fAware}. It is a greedy algorithm that allocates tasks to
workers one by one. Given the task-to-worker bipartite graph $G(E,V)$, s.t. $
V = \mathcal{T} \cup \mathcal{W}$, it considers tasks in the increasing order
of the node in-degree. Tasks with less candidates are placed first, as the
ones with higher degree have more flexibility. For each task, each worker with
remaining capacity higher than $0$ is considered as a candidate and the one
with the lowest \emph{local assignment ratio} is selected. If there are more than one worker with the same \emph{LAR}, the one with
the higher denominator is selected. Recall that
if the capacity of a worker, $w_i.c$, is lower than the number of tasks she has
accepted, we only consider the first $w_i.c$ offers.
The process continues until all tasks are visited. When a task $t_j$
is assigned to a worker $w_i$, the input graph, the total earnings, and the residual
capacity of $w_i$ are updated accordingly. 

Our experimental results confirm the effectiveness of \emph{F-Aware} in terms of
running time, task allocation ratio, and fairness it achieves.

  \section{Online Task Allocation}\label{sec:online}

\noindent While in offline \fb{allocation} based applications, all the tasks and
availabilities are known in advance, in real-time environments, they can
appear at anytime, anywhere~\cite{ref:microTask}. Consider an example scenario, in which
a customer would like to travel to airport and asks the crowd for a ride.
Overnight reply to this request would be too late, as the user would take a
taxi after waiting for a relatively small amount of time. In fact, these
dynamically arriving requests require online processing. Therefore, the
problem of \emph{allocating tasks to workers in a dynamically changing 
environment} raises.

While aforementioned example requires instant response to a customer, there
are also some crowdsourced delivery applications that allow decision maker
system to wait for a period, before allocating the task. Online shopping is an
example of such applications. In this case, a seller can wait to group
deliveries by their destinations and allocate only one worker for multiple
packages. During this waiting period, system collects mini-batches of tasks
and worker availabilities for a period and they are processed against each
other when the period expires. In this section, we present our modifications
on the 2-phase \fb{allocation} model to cover both instant and mini-batch
\fb{allocation} scenarios.

Figure~\ref{fig:workflow} shows the illustration of the workflow for the
2-phase \emph{allocation} model. Our first modification for adapting this workflow to
online task allocation is adding two windows on the selection of nominees.
These windows are for the tasks and for the availabilities, respectively. They
are neither sliding nor tumbling windows. When a task or availability arrives,
it is appended into the corresponding window. An availability is removed from
the window when it expires. A task on the other hand can be removed under two
conditions: \textit{i)} it is assigned to a worker, \textit{ii)} it expires.
However, waiting until expiration of a task before allocation might cause
misses, as all satisfying availabilities might expire meanwhile. Therefore, we
also define a window size, in terms of minutes. For every window expiration,
the tasks are processed against availabilities and nominees are identified.
After this step, batched progressive offer and task allocation steps are used
as they are. There are two corner cases about tasks. First, if a task has no
nominees at the time of the window expiration, it stays in the window and
participates in the following window expirations, until its own expiration.
Second, if the task has nominees but all of them rejected the offer, it again
remains in the window for new nominees to arrive. The same task is never
offered to the same nominee more than once.

There are multiple constraints on the window size decision. First of all, it
should be shorter than the smallest time period of the set of all tasks and
availabilities to guarantee processing. Second, the waiting time should not
exceed reasonable response time of the application. An online shopping
application that needs 2-hours delivery guarantee cannot define the window
size as $3$ hours. For applications that require instant reply,
the window size can be set to $0$. In instant
task allocation, when a task arrives, it is processed against the
availabilities window to identify nominees. If there are not any, it is added
to the tasks window and stays there until its expiration time. When an
availability arrives, all tasks and availabilities present in the window are
processed, since this availability might be completing a partial match. A
partial match is possible when a worker has a satisfying availability for only
one of the steps of the task. We leave the detailed discussion of partial
match processing as future work. In our experimental evaluation, 
we study the feasibility and effectiveness
of our allocation model with various window sizes, including instant
allocation, i.e. window size equals to $0$.

  \section{Experimental Evaluation}\label{sec:experimental}
\noindent In this section, we present the detailed evaluation of our proposed 2-phase
\fb{allocation} model and the fairness-aware task allocation algorithm,
\emph{F-Aware}.

To easily determine the superiority
of a solution over other solutions, we combine optimization goals
into a single parametric objective function $\mathcal{O}$ and 
define it as:

\begin{equation}
\mathcal{O} = TAR \times e^{- (\rho * \mathcal{F})}
\label{eq:obj}
\end{equation}

\rev{Since the goals are maximizing the TAR while minimizing the unfairness,
the objective function is proportional to TAR. But the exponential part of it is 
inversely proportional with the global unfairness metric $\mathcal{F}$.}
To enable the system to prioritize one component of the objective over the other,
we introduce the parameter $\rho$, where $0 \leq \rho \leq 1.$ 
\rev{By setting $\rho$ to $0$, one can simplify this objective to task allocation ratio only. 
Higher values of it will increase the importance of the unfairness in the overall objective.}

The evaluation includes four sets of experiments. In the first set, we compare \emph{F-Aware} with 4 other competitor algorithms in terms of
task allocation ratio~\emph{(TAR)}, \emph{unfairness} and \fb{value of the \emph{objective function}}.
Recall that the objective function, as defined in Eq.~\ref{eq:obj}, is the combined metric of \emph{TAR} and \emph{unfairness}.
The na\"{i}ve approach of assigning tasks to workers is random selection
among the candidates, referred to as \emph{Random} in the performance graphs. 
The second approach is to select \emph{Least Allocated Worker First(LAF)}.
The intuition behind this approach is trying to reserve room for further task allocations.
Existing work of task allocation in spatial crowdsourcing mostly use Nearest Neighbor Priority 
strategy~\cite{ref:kazemi,ref:fairAllocation} to capture the spatial-aspect of the problem.
\cite{ref:kazemi} introduces allocation techniques 
based on location entropy, and~\cite{ref:fairAllocation} extends
nearest worker priority technique with pro-active deployment of 
workers to geo-grids. We only prioritize the nearest worker since we do not make
assumptions on distribution of tasks. This algorithm is referred to as \emph{Nearest} in this section.
Lastly, we use successive shortest paths algorithm which solves the minimum cost flow(MCF) problem~\cite{ref:ssp}.
This algorithm is optimal on Task Allocation Ratio, hence it is used 
to evaluate \emph{TAR} of all algorithms. 
The second set of experiments studies the efficiency of our batch incremental offer strategy, that is how different
values of $\epsilon$ and $\theta$ affect the $k$ value, thus assignment ratio, and \fb{unfairness}.
The third set studies \emph{online task allocation}, presenting task allocation ratio and
unfairness as a function of window size. Finally, the last set is the
sensitivity study that presents task allocation ratio and unfairness of different
time period lengths as a function of coefficient of mean. As we will detail soon, length of
the time period and coefficient of the mean are two variables we use to adapt real-world data to our setup.

We implemented all algorithms using Java 1.8.
All experiments were executed on a Linux server with
2 Intel Xeon E5520 2.27GHz CPUs and 64GB of RAM. 

\begin{figure*}[ht]
	\begin{subfigure}{0.24\linewidth}
		\includegraphics[width=\linewidth]{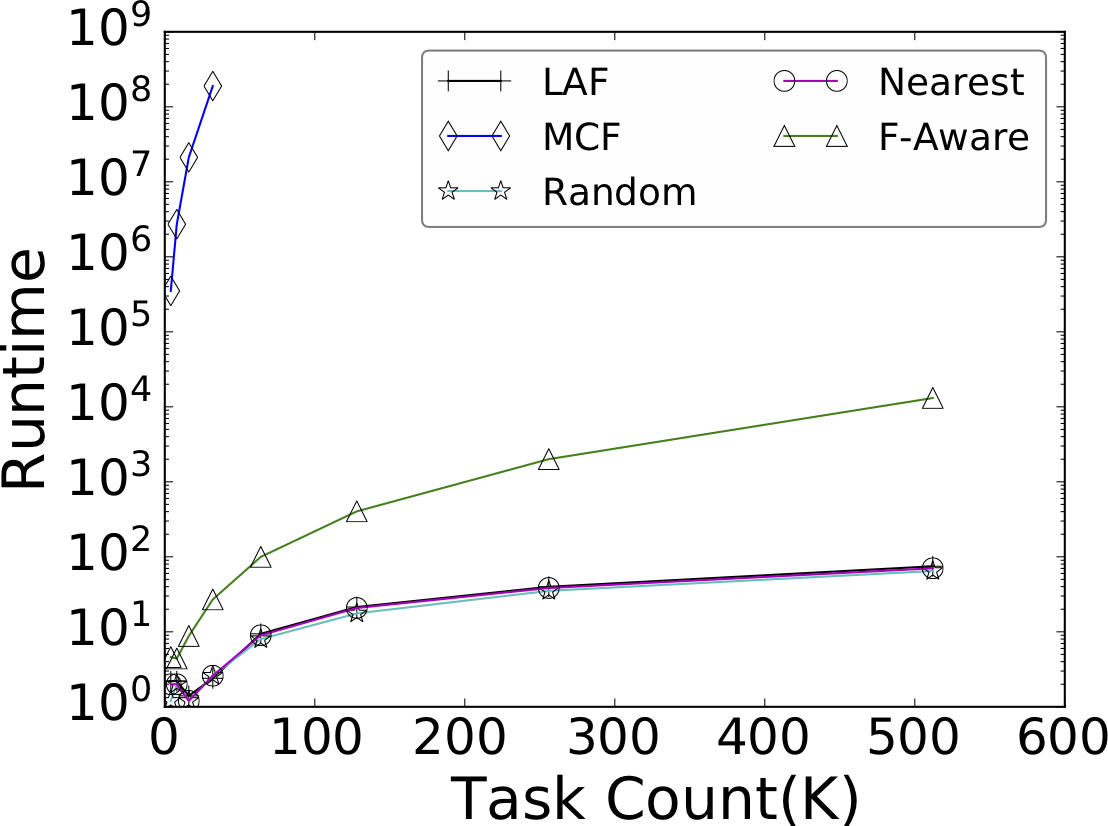}
 		\caption{Runtime}
		\label{fig:scalability_runtime}
	\end{subfigure}
	\hspace{0.0001\linewidth}
  	\begin{subfigure}{0.24\linewidth}
		\includegraphics[width=\linewidth]{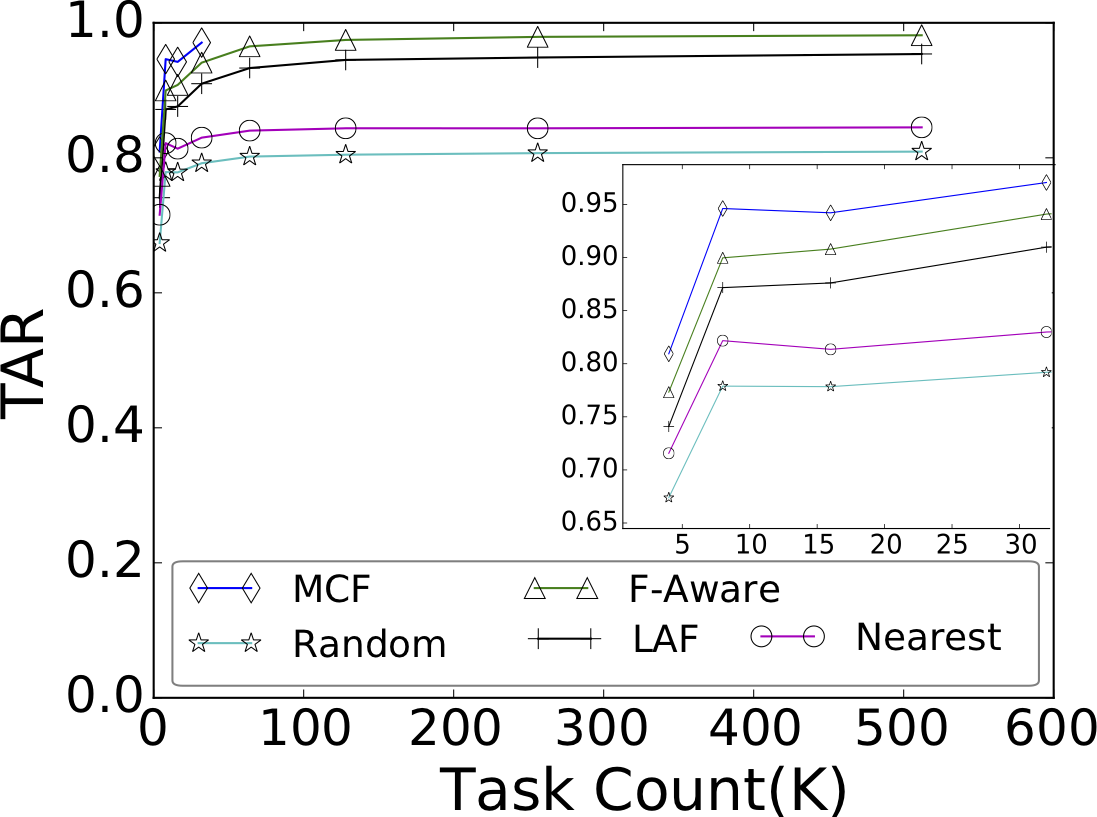}
 		\caption{TAR}
		\label{fig:scalability_tcr}
	\end{subfigure}
  	\hspace{0.0001\linewidth}
	\begin{subfigure}{0.24\linewidth}
		\includegraphics[width=\linewidth]{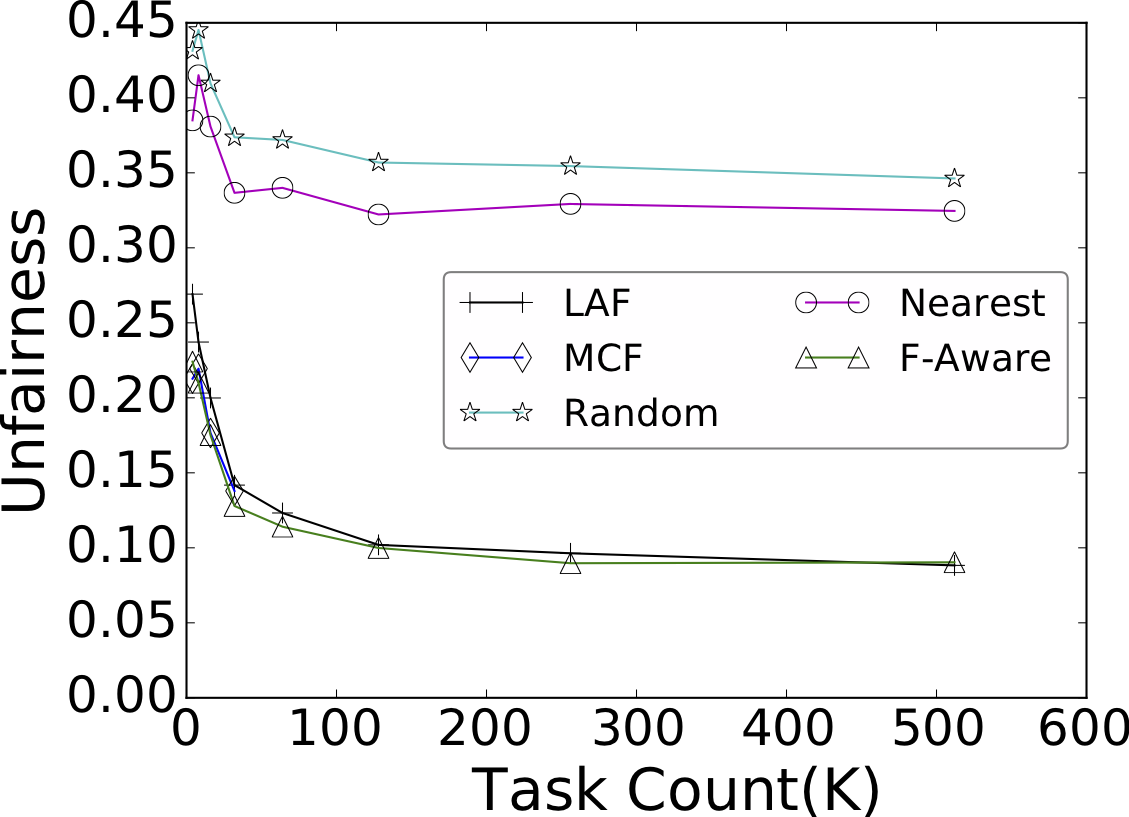}
  		\caption{Unfairness}
  		\label{fig:scalability_unfairness}
	\end{subfigure}
	\begin{subfigure}{0.24\linewidth}
		\includegraphics[width=\linewidth]{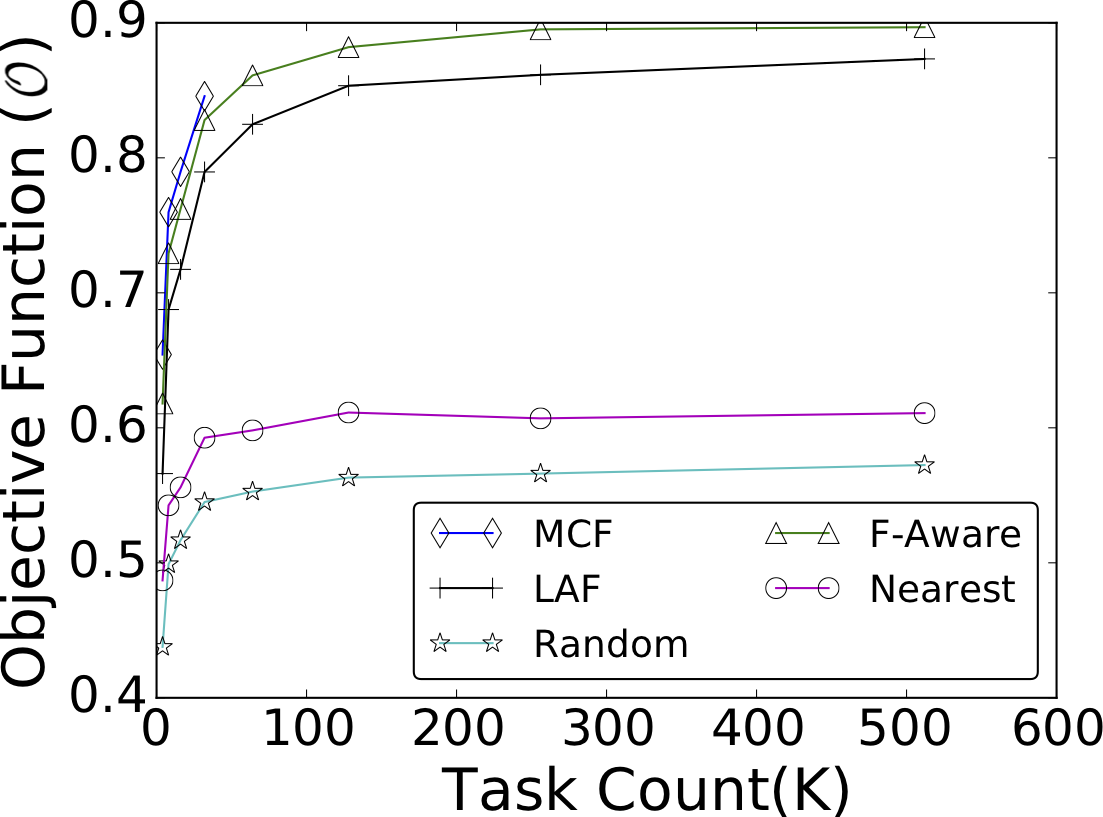}
 		\caption{Objective}
 		\label{fig:scalability_objective}
	\end{subfigure}
	\caption{Scalability}
	\label{fig:scalability}
\end{figure*}
\begin{figure*}
	\hspace{0.0001\linewidth}
	\begin{subfigure}{0.24\linewidth}
  		\includegraphics[width=\linewidth]{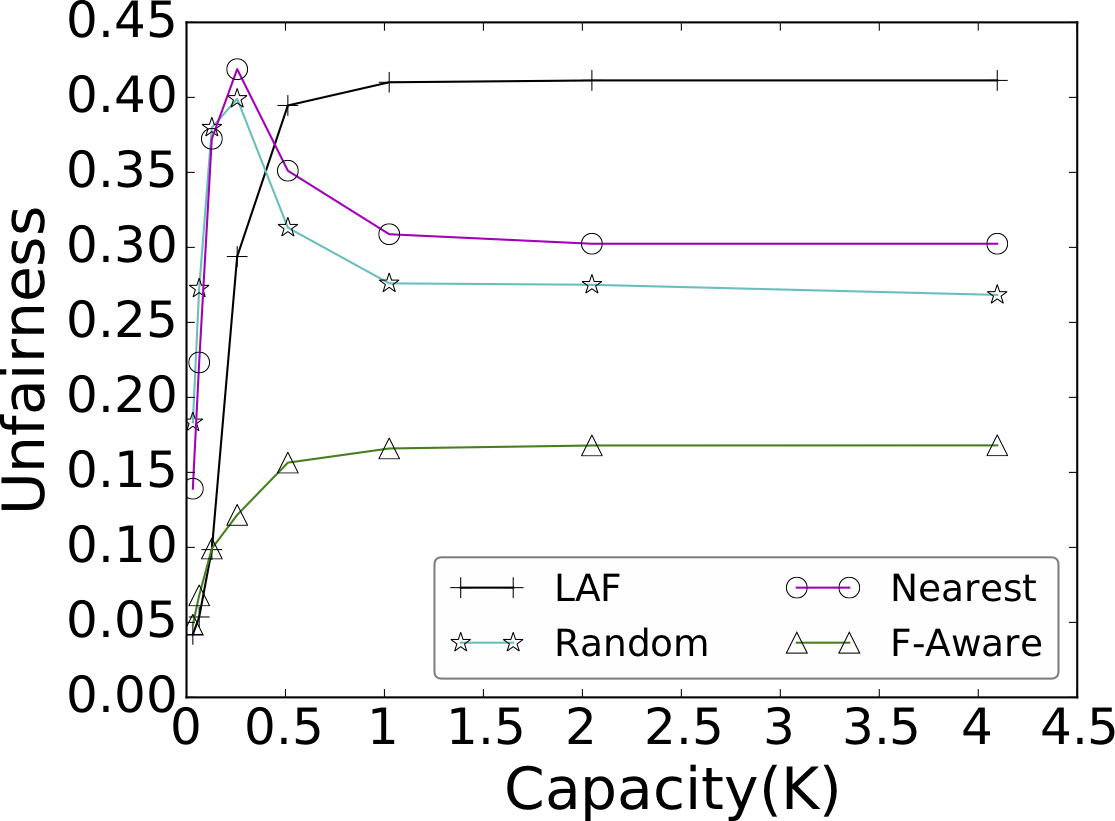}
  		\caption{Unfairness}
		\label{fig:capacity_unfairness}
	\end{subfigure}
	\hspace{0.0001\linewidth}
	\begin{subfigure}{0.24\linewidth}
  		\includegraphics[width=\linewidth]{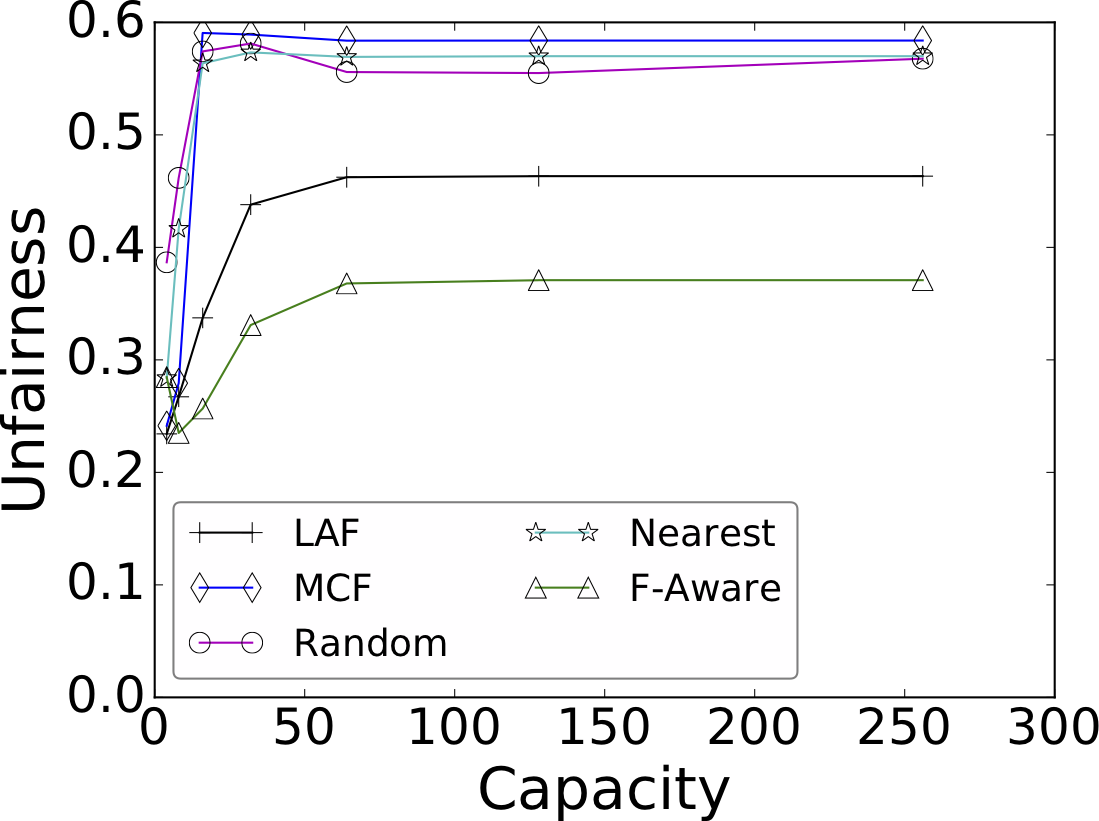}
  		\caption{Unfairness-small task set}
		\label{fig:capacity_small_unfairness}
	\end{subfigure}
  	\hspace{0.0001\linewidth}
	\begin{subfigure}{0.24\linewidth}
  		\includegraphics[width=\linewidth]{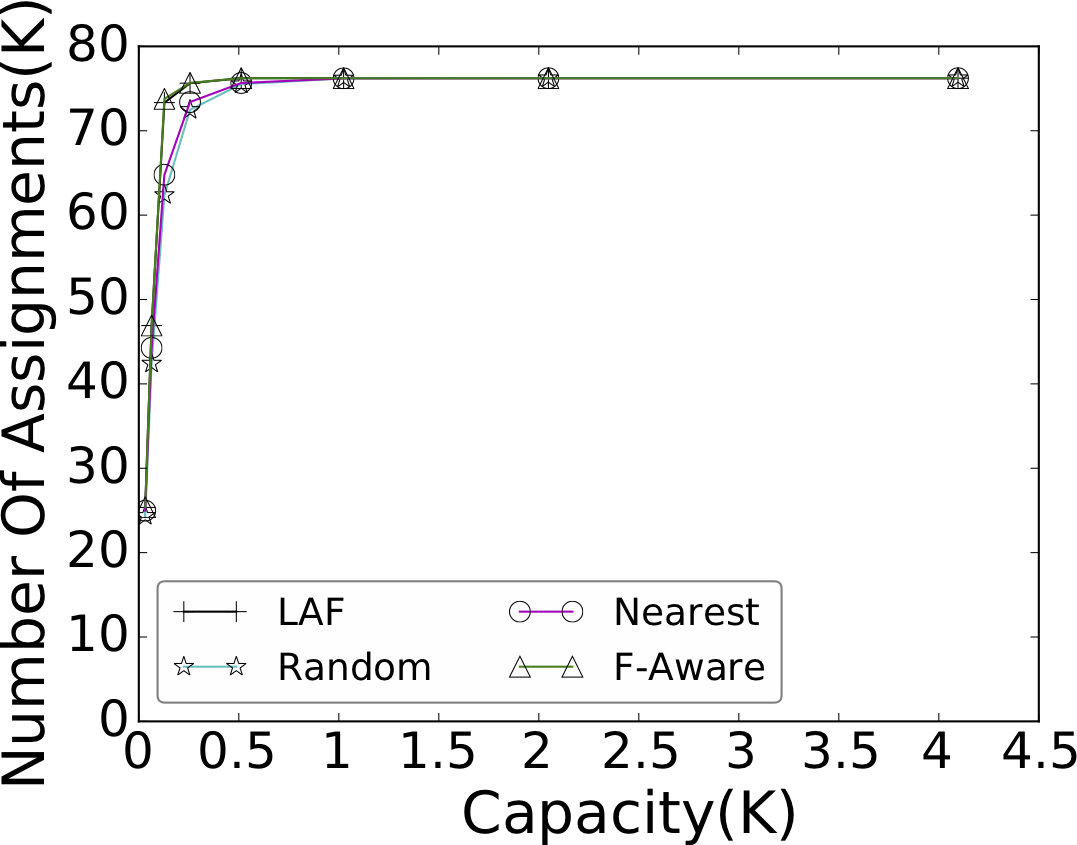}
  		\caption{\# Of Completed Tasks}
		\label{fig:capacity_nct}
	\end{subfigure}
  	\hspace{0.0001\linewidth}
	\begin{subfigure}{0.24\linewidth}
		\includegraphics[width=\linewidth]{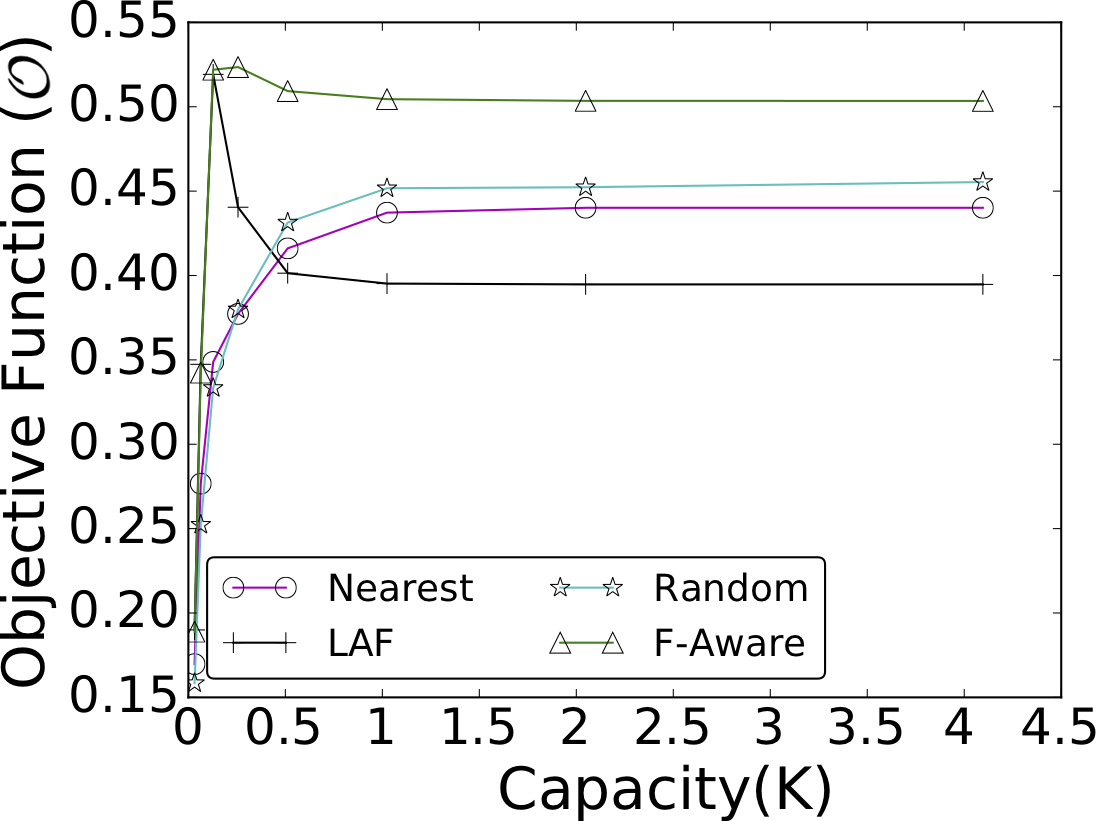}
 		\caption{Objective}
		\label{fig:capacity_objective}
	\end{subfigure}
	\caption{Effect of the Worker Capacity}
	\label{fig:capacity}
\end{figure*}

\textbf{Dataset.} Experiments are performed using two real-world datasets. The first dataset
contains the Foursquare check-ins from New York City for the month of May
2012~\cite{ref:4sqData}\footnote{sites.google.com/site/yangdingqi/home/foursquare-dataset}. 
This dataset contains around $50{,}000$ check-ins from
$987$ different users. The second dataset is a taxi-trip dataset for Manhattan,
for the same time period\footnote{www.nyc.gov/html/tlc/html/about/trip\_record\_data.shtml}. 
We used up to $512{,}000$ randomly sampled rows from
the taxi dataset. Half million tasks for a city, for one month, is a fair workload,
considering assignment could be done independently for each city.
Each row of the check-in data contains a user id, time of
the check-in, and the location of it. Each row in the taxi dataset contains
the time and the location of the pick-up and the drop-off. It also contains
the cost of the trip. To simulate crowdsourced delivery behavior, we use taxi trips as
tasks and check-ins as spatio-temporal availabilities of the workers. For both datasets, we
converted all time points to time ranges by adding time periods. \fb{Given a point in time, $p$, a
period of length $\Delta T$ is created with beginning and end points $[p, p+(\Delta T)]$.}
\fb{We treat pick-up location and time of a taxi trip as the task's source location and source validity time. 
The same applies to drop-off location and time and tasks deliver location and validity.} 
\fb{For each check-in, we use the location of it as the location of the availability and converted
the time of it to a time period as described above.} 
Radius of each availability is sampled from a Normal distribution with mean and standard
deviation calculated using taxi trips. We study the
effect of this conversion as part of our last experiment presented in this
section, by varying the mean size of the time period. For all other
experiments, the mean value for the time period is taken as 2 hours. 
Last, we used the cost of the trip as the reward of the task.

\subsection{Scalability}
In this set, we present two subsets of experiments. In the first subset,
we observe the running time performance, task allocation ratio, and unfairness
as a function of the number of tasks. The number of tasks is doubled for every
data point ranging from $4000$ tasks up to $512{,}000$. Capacities of the
workers are assigned using a Normal distribution and the mean is set to the number of tasks 
over the number of workers. 
\fb{Standard deviation of the distribution is set to mean over $4$ to make sure all capacities are at least $0$.} 
Using this tight capacity assignment for this
particular experiment set, we ensure that the capacity is barely enough for
allocating all the tasks. This gives a clear advantage to algorithms that can
allocate tasks close to optimal. In the second subset, we observe the task allocation
ratio and unfairness as a function of worker capacity. $128{,}000$ tasks are used, and 
capacities are doubled for every data point ranging from $32$ to $4096$.
To observe the behavior of the \emph{MCF} algorithm, we also present unfairness as
a function of capacity using $8000$ tasks~(Figure~\ref{fig:capacity_small_unfairness}).
In this experiment, capacities are ranging from $4$ to $256$.

Figures~\ref{fig:scalability}, and~\ref{fig:capacity} present our scalability related results. 
In all figures, the $x$-axis represents either the number of tasks to allocate, or 
the capacity of workers, and $y$-axis
represents the performance metric. Different series represent different
assignment algorithms. Figure~\ref{fig:scalability_runtime} plots the running
time as a function of the number of the tasks. We make two observations from
the figure. First, and most importantly, \emph{F-Aware} runs $10^7$ times
faster than the Minimum Cost Flow \emph{MCF} algorithm. For $32{,}000$ tasks, the
running time of the MCF is $1.89\times10^8$ milliseconds while \emph{F-Aware} completes
in $27$ milliseconds. Because of the long evaluation time, we do not present
\emph{MCF} results for more than $32,000$ tasks. Secondly, the running time of the
\emph{F-Aware} algorithm is linear with the task count. Increasing from
$4000$ tasks to $512{,}000$ tasks, the running time increases from $4.6$
milliseconds to $13{,}130$ milliseconds. The difference between the running times of
\emph{Random}, \emph{LAF} and \emph{Nearest} assignment algorithms are
negligible.

Figure~\ref{fig:scalability_tcr} plots the task allocation ratio as a function
of the number of tasks. To increase readability, it also includes the zoomed small figure 
of data points between $[4000, 32{,}000]$.
Since any allocation algorithm will left tasks with no candidates
unassigned, in this experiment we consider only tasks with at least one candidate. 
We observe that all allocation algorithms are able to hold their allocation ratio with the
increasing number of tasks. The most important observation is that
\emph{F-Aware} is able to assign $96.9\%$ of the tasks that are allocated by
\emph{MCF}. For $32,000$ tasks, \emph{MCF} reaches to $97\%$ task allocation ration, while
\emph{F-Aware} allocates $94\%$ of all tasks. When we double the number of
tasks, \emph{F-Aware} still allocates $96.5\%$ of tasks, while \emph{Random} and \emph{Nearest}
worker allocation algorithms stay at $80\%$ and $84\%$, respectively.
\fb{In terms of \emph{TAR}, \emph{LAF} is the best competitor of \emph{F-Aware}. This is expected
as its goal is to increase the number of allocated tasks.} However, \emph{TAR} of
\emph{F-Aware} is still higher. For $512{,}000$ tasks, \emph{F-Aware} 
reaches to $98.1\%$ task allocation ratio, while \emph{LAF} reaches to $95.3\%$.

We discuss the Figures~\ref{fig:scalability_unfairness}, ~\ref{fig:capacity_unfairness}, 
~\ref{fig:capacity_small_unfairness} and~\ref{fig:capacity_nct} together as they are complementary.
First three figures plot the unfairness as a function of task count, capacity and capacity
respectively. The last one represents the number of assigned tasks, as a function of capacity.
In Figure~\ref{fig:capacity_small_unfairness}, we use only $8000$ tasks to include \emph{MCF}
algorithm. In the first Figure, (\ref{fig:scalability_unfairness}), we use up to $512{,}000$
tasks but exclude \emph{MCF} after $32{,}000$ tasks, because of the impractically long running
time. For the remaining two, we use $128{,}000$ tasks, and capacity of the workers varies from $32$ to
$4096$. 

At this point, it would be useful to recall, \emph{LAR} of a worker is the ratio of
revenue she made from completed task over total reward of the offers she has accepted.
However, since capacity limits the assignable task count, if a worker $w_i$ accepts more offer
than her capacity $w_i^c$, we only consider first $w_i^c$ offers when calculating \emph{LAR}.
Therefore, when evaluating performance of an algorithm in terms of fairness, one must consider
two cases when capacity limits assignments, and when there are more than enough room for assignments.

When worker capacities are enough to serve all tasks, we observe a significant 
difference between unfairness values. In Figure~\ref{fig:capacity_small_unfairness}, 
for small dataset, \emph{Random}, \emph{Nearest}, and \emph{MCF} 
perform similar. In contrast, unfairness metric of \emph{LAF} $0.8\times$ of the same metric of those three.
We observe \emph{F-Aware}, performs best among all.
Unfairness metric \emph{MCF} algorithm is $1.5\times$ that of our \emph{F-Aware} algorithm
when capacity is $256$.
For the larger task set,
the difference between \emph{LAF} and \emph{F-Aware} becomes even more significant (figures~\ref{fig:capacity_unfairness},~\ref{fig:capacity_nct}).
When the capacity is set to $1024$ unfairness value of \emph{LAF} is $0.41$ while \emph{F-Aware}
has only $0.16$ unfairness. For same data point \emph{Nearest}, and \emph{Random} have
$0.30$, $0.27$ unfairness values respectively. We make two additional observations from this 
figure. First, up to $256$ capacity, unfairness values increase. This is because 
capacity of workers less than number of acceptances. 
In this set of experiments each worker accepts $215$ offers in average. 
After this point number of accepted tasks used when calculating \emph{LAR}.
Using Figure~\ref{fig:capacity_nct} we observe all algorithms reach maximum
number of allocated tasks at capacity $1024$, which reflected as stabilized
unfairness values in Figure~\ref{fig:capacity_unfairness}.
Second, \emph{LAF} performs poorer than \emph{Nearest} and \emph{Random} assignment approaches.
Since it does not take user input into account, (i.e. accepting offers)
when tasks are distributed evenly, workers who have accepted small number of offers
have $LAR = 1$, while workers with large number of acceptances have too low
\emph{LAR} values.

When capacity is set to a too low value, we cannot observe significant difference between unfairness 
values of \emph{MCF}, \emph{LAF}, and \emph{F-Aware}.
This is because all workers are fully allocated.
On the other hand, in Figure~\ref{fig:scalability_unfairness} values of unfairness metric
for \emph{Random}, and \emph{Nearest} is around $3.6\times$ that of \emph{F-Aware}. The reason
behind is system could serve less number of tasks (Figure~\ref{fig:scalability_tcr}), when
one of these two assignment algorithm is used. 

Figures~\ref{fig:scalability_objective} and~\ref{fig:capacity_objective} present the performance
in terms of our objective, Eq~\ref{eq:obj}, as a function of task count and capacity respectively. 
In these experiments, the $\rho$ parameter is set to $1$, to observe the balanced outcome of \emph{task allocation ratio} 
and \emph{fairness}.
In Figure~\ref{fig:scalability_objective}, we observe that the 
difference between the objective values of \emph{F-Aware} and \emph{MCF} are negligible.
\emph{F-Aware} performs
as good as \emph{MCF}, in terms of our objective, with $10^7$ times faster processing speed. Another observation
from the same figure is that, \emph{F-Aware} performs $4\%$ better than \emph{LAF}, even with limited capacity. For
$128,000$ tasks, \emph{F-Aware} has $0.86$ objective value while \emph{LAF} has $0.82$. When there are more than
enough room for assignments, we observe that the gap between \emph{F-Aware} and \emph{LAF} becomes even more significant,
as shown by Figure~\ref{fig:capacity_objective}. For a capacity value of $4096$ the objective value of \emph{F-Aware} is $0.5$, while
\emph{LAF} could reach only $0.39$. Objective values of \emph{Nearest} and \emph{Random} are better than
\emph{LAF}, because of the lower unfairness values, but still they are far from performing as good as \emph{F-Aware}.

To summarize this experiment set, one can say that using \emph{MCF} is impractical due to its long
running time. In contrast, \emph{F-Aware} runs $10^7$ times faster. When only \emph{TAR}
is considered, \emph{LAF} performs similar to \emph{F-Aware}, but the other two approaches,
\emph{Random} and \emph{Nearest}, leave $19\%$ and $15\%$ of all tasks unassigned, respectively.
While \emph{LAF} is the best competitor of \emph{F-Aware}, in terms of \emph{TAR} and runtime,
its unfairness metric value is $2.5\times$ that of \emph{F-Aware} and its objective value is $80\%$ 
that of our \emph{F-Aware} algorithm.

\subsection{Effect of the Batch Size}
\begin{figure*}[ht]
	\begin{subfigure}{0.24\linewidth}
		\includegraphics[width=\linewidth]{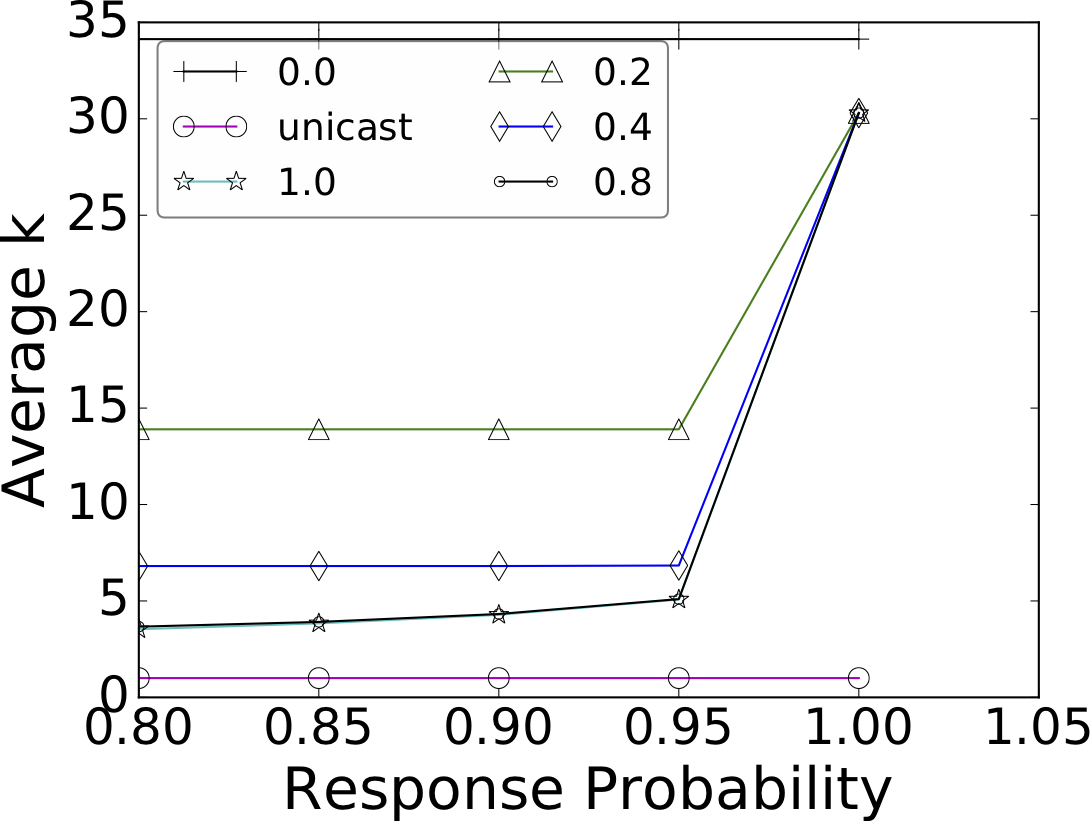}
 		\caption{Average K}
		\label{fig:batch_avgK}
	\end{subfigure}
	\hspace{0.0001\linewidth}
  	\begin{subfigure}{0.24\linewidth}
		\includegraphics[width=\linewidth]{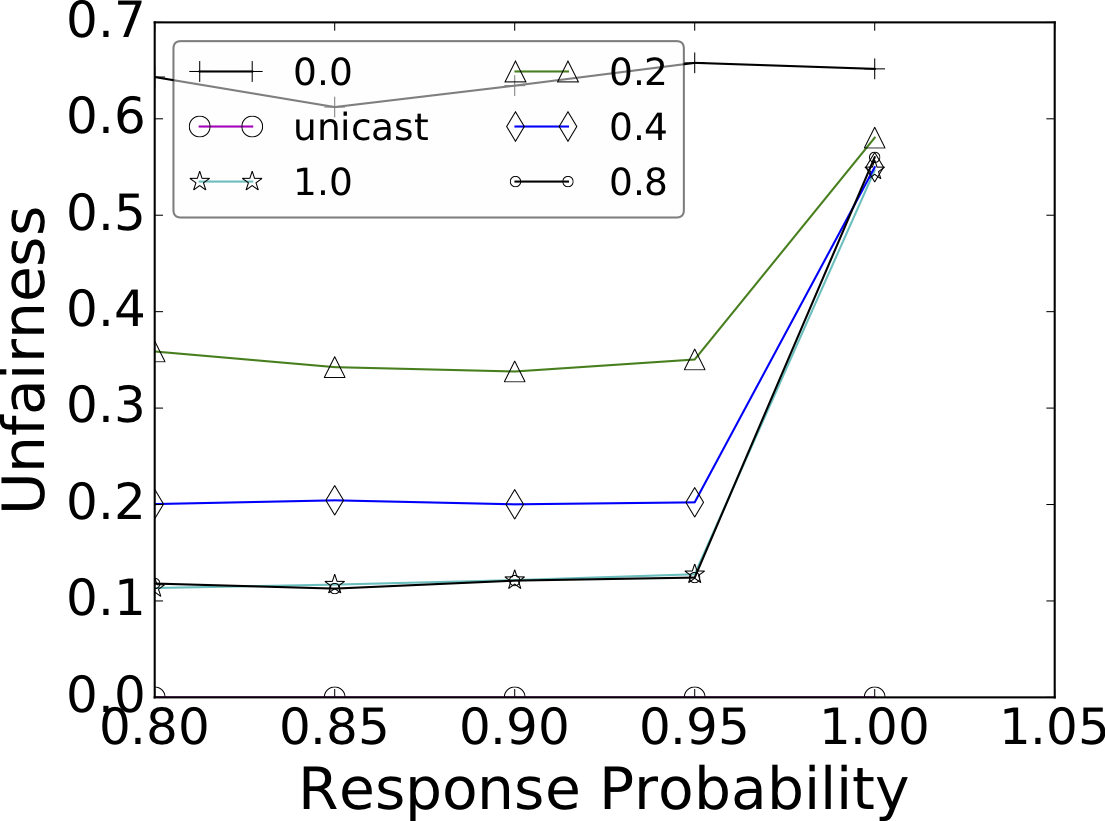}
 		\caption{Unfairness}
		\label{fig:batch_unfairness}
	\end{subfigure}
  	\hspace{0.0001\linewidth}
	\begin{subfigure}{0.24\linewidth}
  		\includegraphics[width=\linewidth]{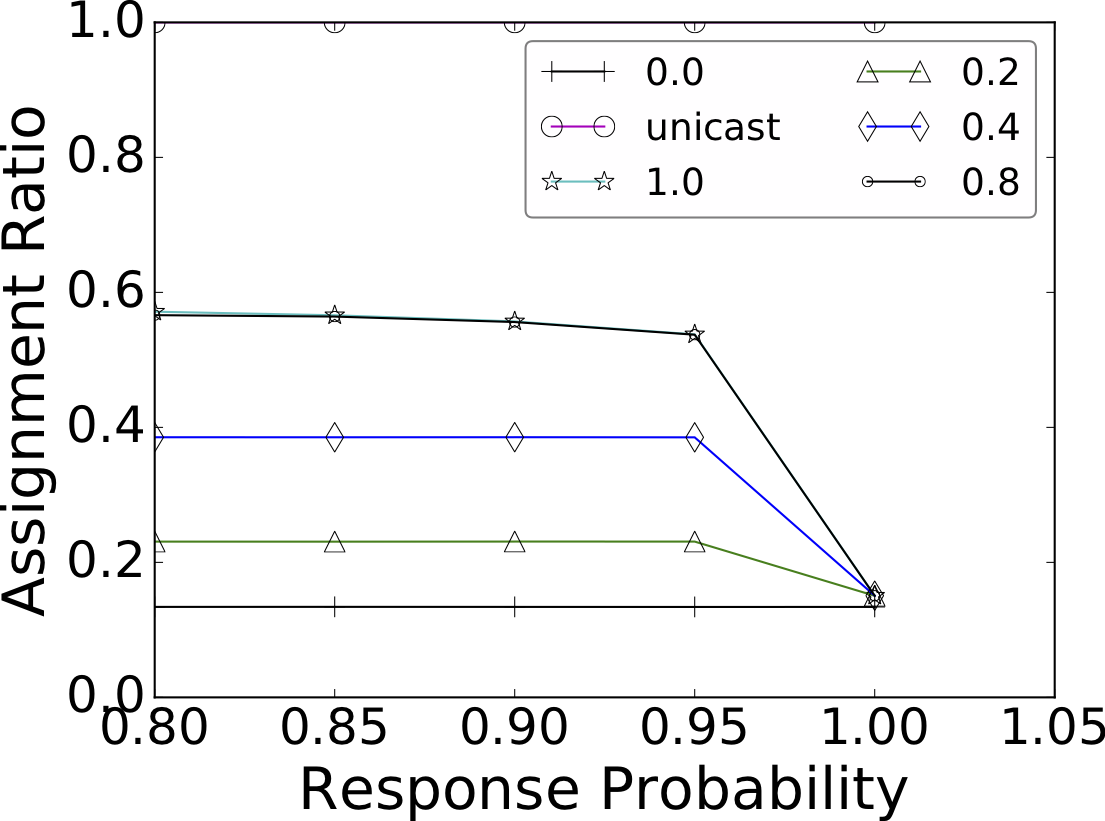}
  		\caption{Assignment Ratio}
		\label{fig:batch_ar}
	\end{subfigure}
	\hspace{0.0001\linewidth}
	\begin{subfigure}{0.24\linewidth}
  		\includegraphics[width=\linewidth]{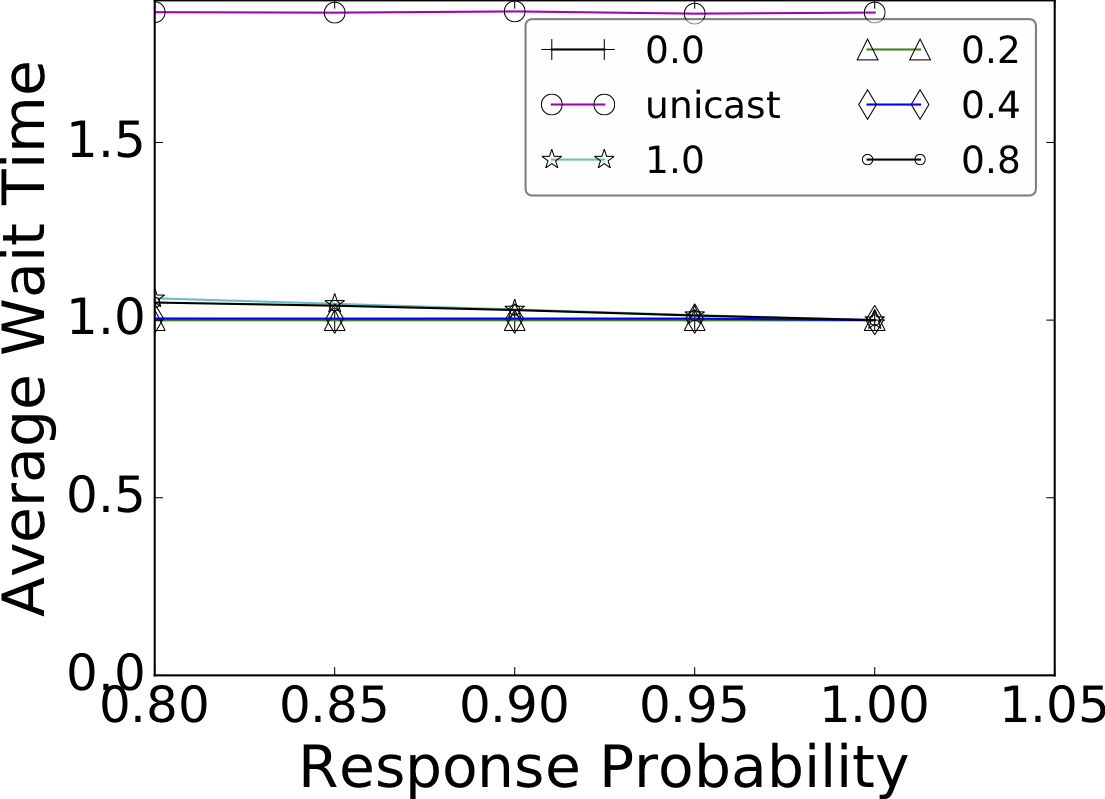}
  		\caption{Average Wait Time}
		\label{fig:batch_wait}
	\end{subfigure}
	\caption{Effect of the Batch Size}
	\label{fig:batch}
\end{figure*}
\begin{figure*}[ht]
	\begin{subfigure}{0.24\linewidth}
		\includegraphics[width=\linewidth]{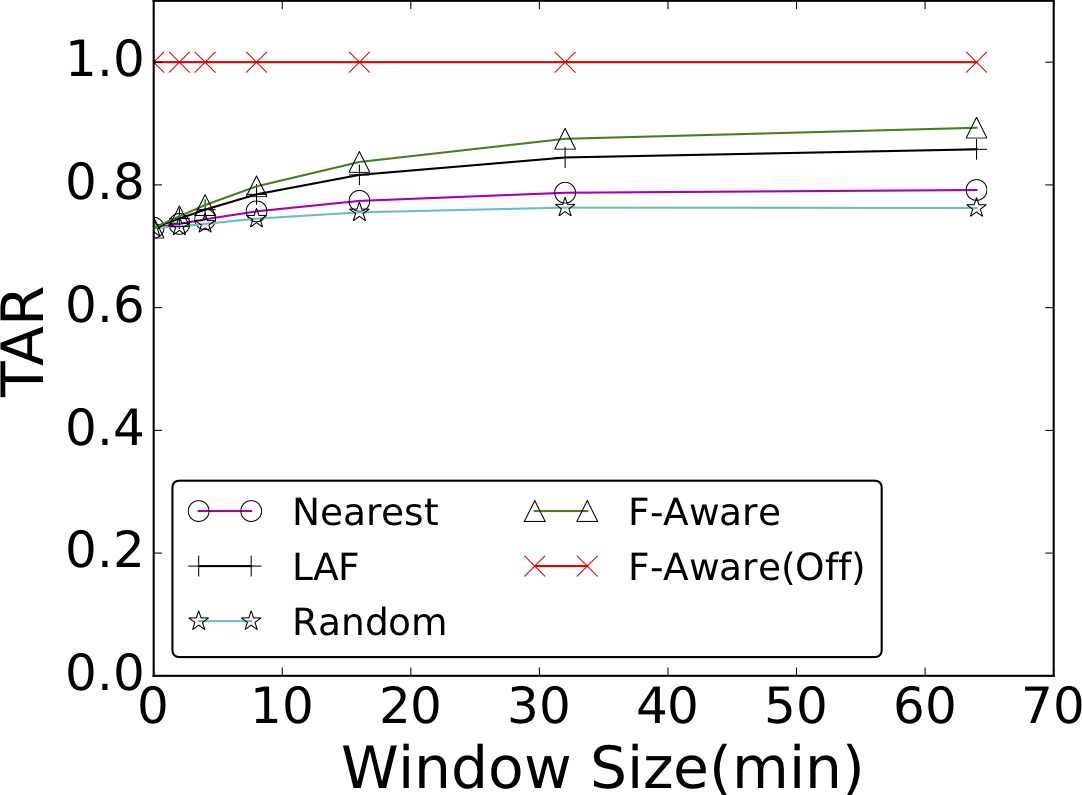}
 		\caption{TAR Rel. to Offline}
		\label{fig:online_tcr}
	\end{subfigure}
	\hspace{0.0001\linewidth}
  	\begin{subfigure}{0.24\linewidth}
		\includegraphics[width=\linewidth]{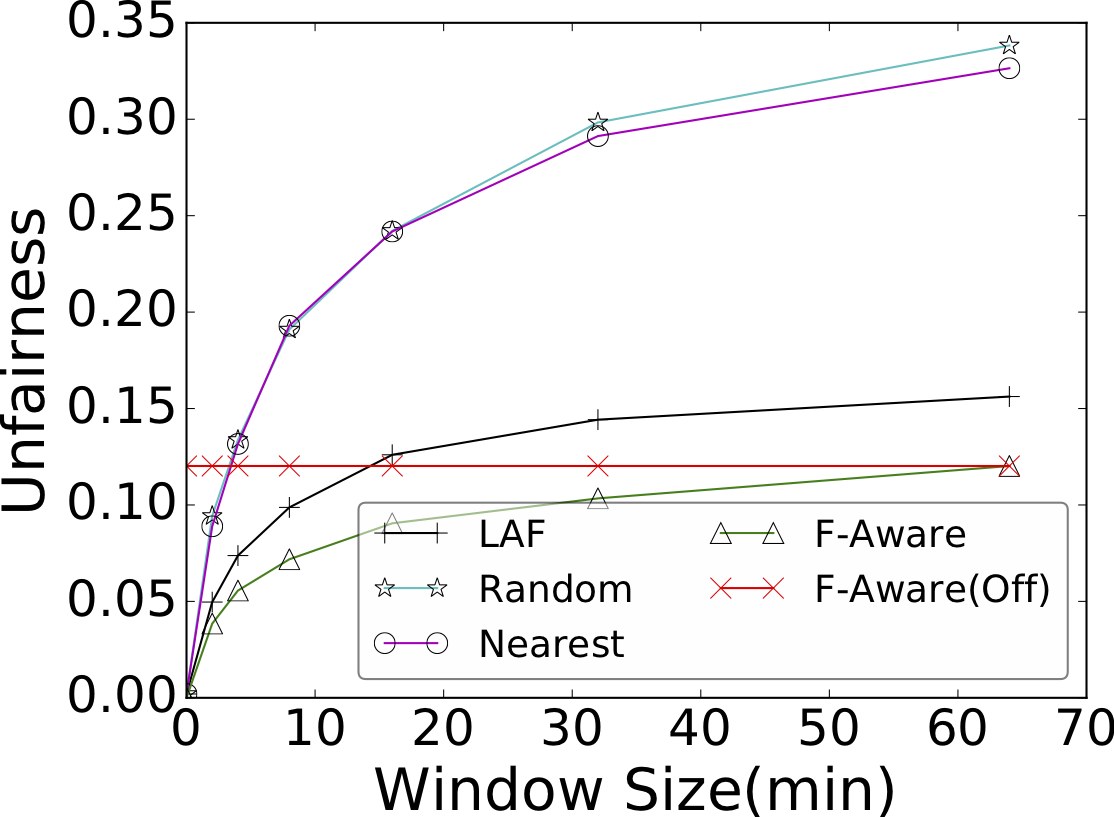}
 		\caption{Unfairness Rel. to Offline}
		\label{fig:online_unfairness}
	\end{subfigure}
  	\hspace{0.0001\linewidth}
	\begin{subfigure}{0.24\linewidth}
  		\includegraphics[width=\linewidth]{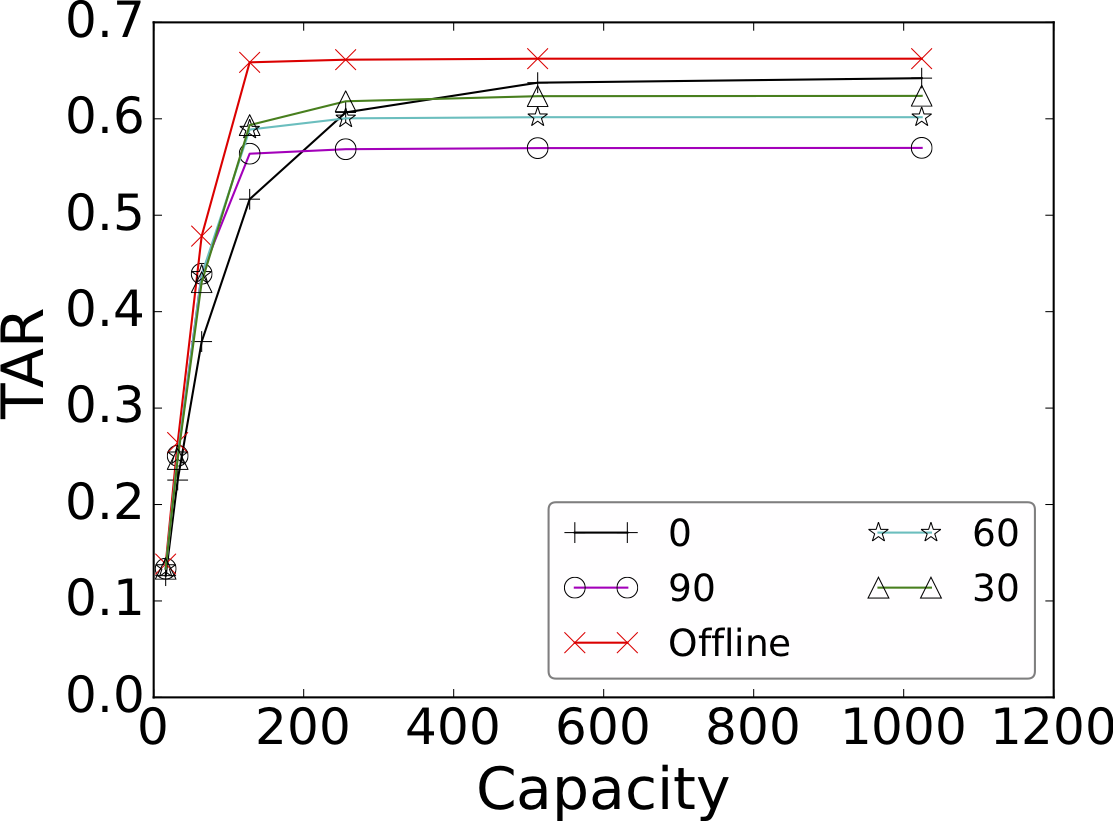}
  		\caption{TAR}
		\label{fig:online_cap_tcr}
	\end{subfigure}
	\hspace{0.0001\linewidth}
	\begin{subfigure}{0.24\linewidth}
  		\includegraphics[width=\linewidth]{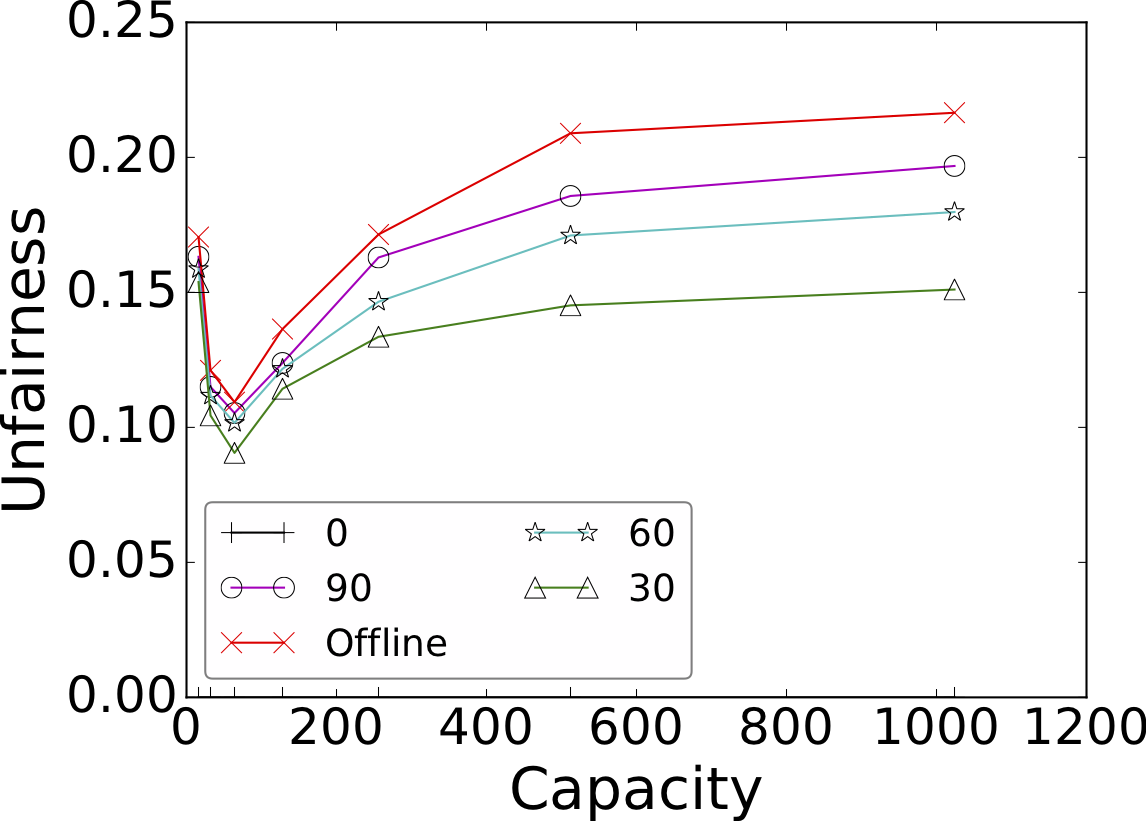}
  		\caption{Unfairness}
		\label{fig:online_cap_unfairness}
	\end{subfigure}
	\caption{Online Task Allocation}
	\label{fig:online}
\end{figure*}
We observe average $k$, unfairness, assignment ratio, and average wait time as
a function of the probability of response threshold $\epsilon$.
Figure~\ref{fig:batch} presents our batch size related results. In all figures
$x$-axis represents the value of $\epsilon$ and $y$-axis represents a
performance metric. Different series represent different assignment ratio
thresholds, $\theta$. For all series we use the \emph{F-Aware} assignment
algorithm. For this set of experiments capacity of workers is high enough to
prevent a bottleneck.

Figure~\ref{fig:batch_avgK} plots the change in the value of $k$ as a function
of $\epsilon$. The series with circle marker is the unicast line, i.e. $k$ is
set to $1$. Recall that the value of $k$ is bounded from below by a function
of $\epsilon$ and bounded from above by a function of $\theta$. We select
the largest $k$ inside this range. Higher values of $\epsilon$ and $\theta$
imply tighter bounds. Since $\theta = 0.0$ means unlimited upper bound,
practically it is the broadcasting line. We observe that as long as $\epsilon
\leq 0.95$ the value of $k$ is limited by the assignment ratio threshold
$\theta$. Increasing $\theta$ from $0.2$ to $0.4$ decreases the average
$k$ value from $14$ to $6$. Since task completion is our primary goal, when
lower bound is higher than the upper bound (possible in some cases based on
the definition of Equation~\ref{eq:kequality}), we use the lower bound for
deciding the $k$ value. One can observe this behavior when $\epsilon \geq
0.95$, as all the average $k$ values are closer to broadcasting.

Figure~\ref{fig:batch_unfairness} plots the unfairness as a function of
$\epsilon$. We observe that smaller $k$ values provides a more fair systems.
The unfairness of unicasting is $0$, as whenever  a worker accepts a task,
she will be assigned to it. Whereas the unfairness of broadcasting
is $0.36$. Most importantly, unfairness of multicasting the offer to an
average of $5.8$ nominees is $0.20$. There are two reasons behind this
observation. First, \emph{local assignment ratio} of a worker is negatively
correlated with the number of her acceptances. Therefore, the mean of the set
of \emph{LAR} values increases, which leads to a decrease in the coefficient of
variation metric. Second, and more importantly, when a worker accepts an
offer, probability of her getting the job is higher with the smaller values of
$k$. In the extreme case, that is unicasting, acceptance implies assignment,
hence unfairness = $0$.
We can also observe a similar behavior in Figure~\ref{fig:batch_ar}.
Assignment ratio of the unicasting equals to $1$. For the other series, we can
see that assignment ratio is negatively correlated with the average $k$ value.
For broadcasting, it decreases up to $0.13$. When the average $k$ is $5.8$,
the assignment ratio is $0.38$. Another important observation is theoretically 
$\theta = 1.0$ should be the unicast line but it has $0.6$ assignment ratio.
The reason behind is $\epsilon$ is a stronger constraint than $\theta$.
We do not observe same behavior lines other than $\theta = 1.0$, 
and $\theta = 0.8$

Figure~\ref{fig:batch_wait} plots the average wait time as a function of
$\epsilon$. $y$-axis shows the average number of rounds passed until there is
at least one \emph{candidate}. Recall that the task is offered to nominees in
batches until there is at least one response. Between each round, the system
waits for a predefined period to let nominees decide. For some scenarios, e.g.
on-demand transportation, customers expect almost instant reply. Higher number
of rounds before acceptance leads to late notification to a customer. In case
of broadcasting, since all nominees are notified at once, number of rounds for
response is always $1$. However, as we have just seen, broadcasting leads to
low assignment ratio and high unfairness. At the other extreme, that is
unicasting, the average wait time is $1.8\times$ of broadcasting. Multicasting
is better than both approaches. Multicasting the task to an average of
$5.08$~($\epsilon = 0.8, \theta = 0.95$) nominees leads to only $1.01$
average rounds, while providing only $0.13$ unfairness. 

In summary, one can say that multicasting is better than broadcasting in terms
of \emph{assignment ratio} and \emph{fairness}. Moreover, it beats unicasting
when \emph{average wait time} is considered, while being almost as fair.

\begin{figure*}
	\begin{minipage}{0.49\linewidth}
		\begin{subfigure}{0.49\linewidth}
			\includegraphics[width=\linewidth]{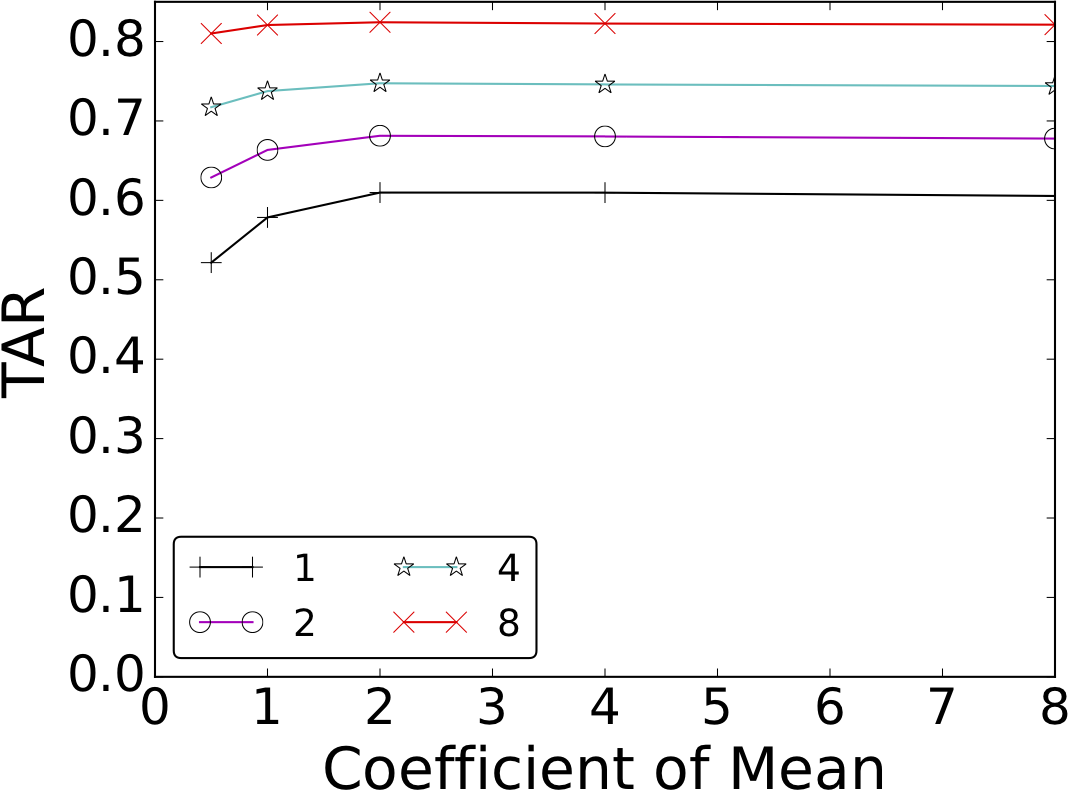}
 			\caption{Task Allocation Ratio}
			\label{fig:sensitivity_tcr}
		\end{subfigure}
		\hspace{0.0001\linewidth}
  		\begin{subfigure}{0.49\linewidth}
			\includegraphics[width=\linewidth]{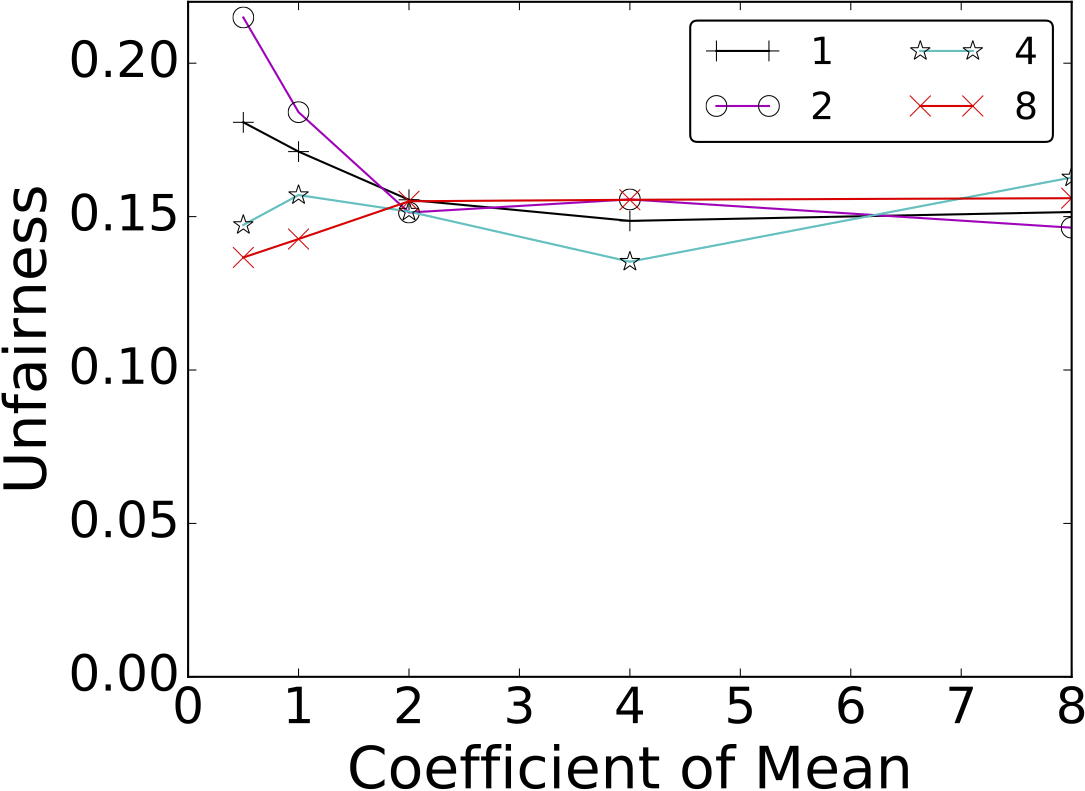}
 			\caption{Unfairness}
			\label{fig:sensitivity_unfairness}
		\end{subfigure}
		\caption{Sensitivity Experiment}
		\label{fig:parameters}
	\end{minipage}
	\hspace{0.0001\linewidth}
	\begin{minipage}{0.49\linewidth}
		\begin{subfigure}{0.49\linewidth}
			\includegraphics[width=\linewidth]{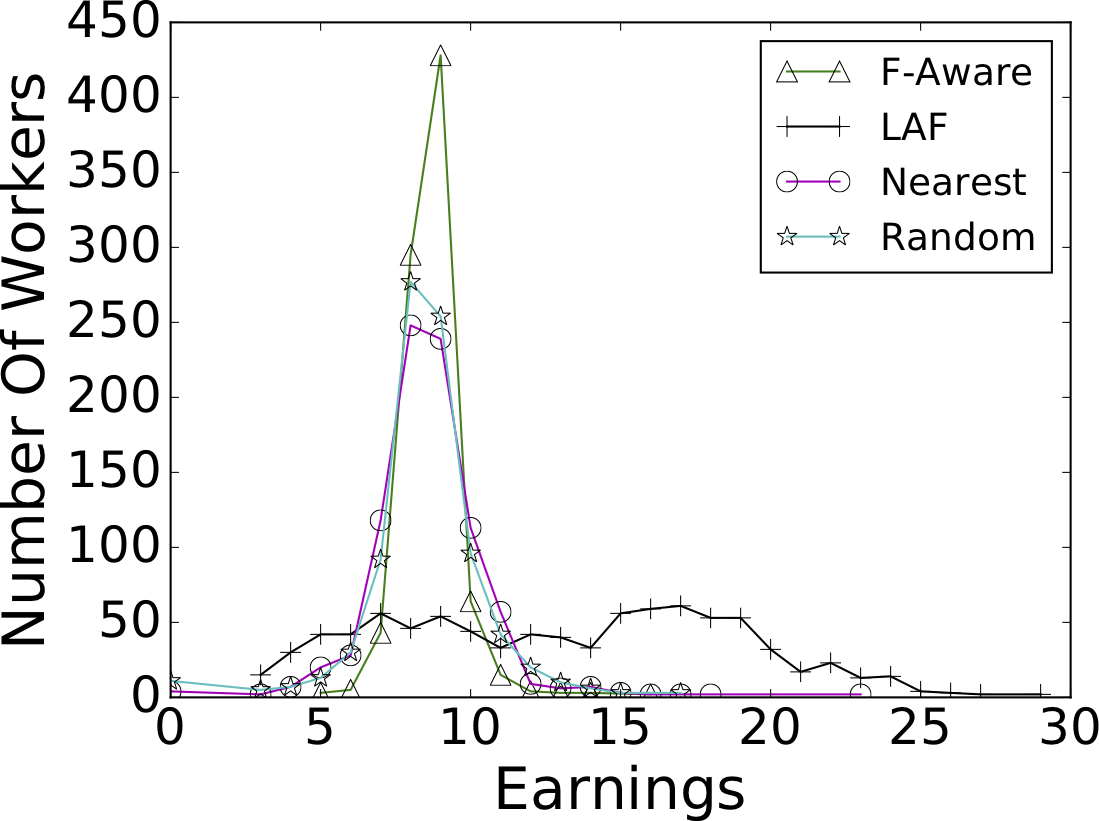}
 			\caption{Earning Distribution}
			\label{fig:discussion_dist}
		\end{subfigure}
		\hspace{0.0001\linewidth}
  		\begin{subfigure}{0.49\linewidth}
			\includegraphics[width=\linewidth]{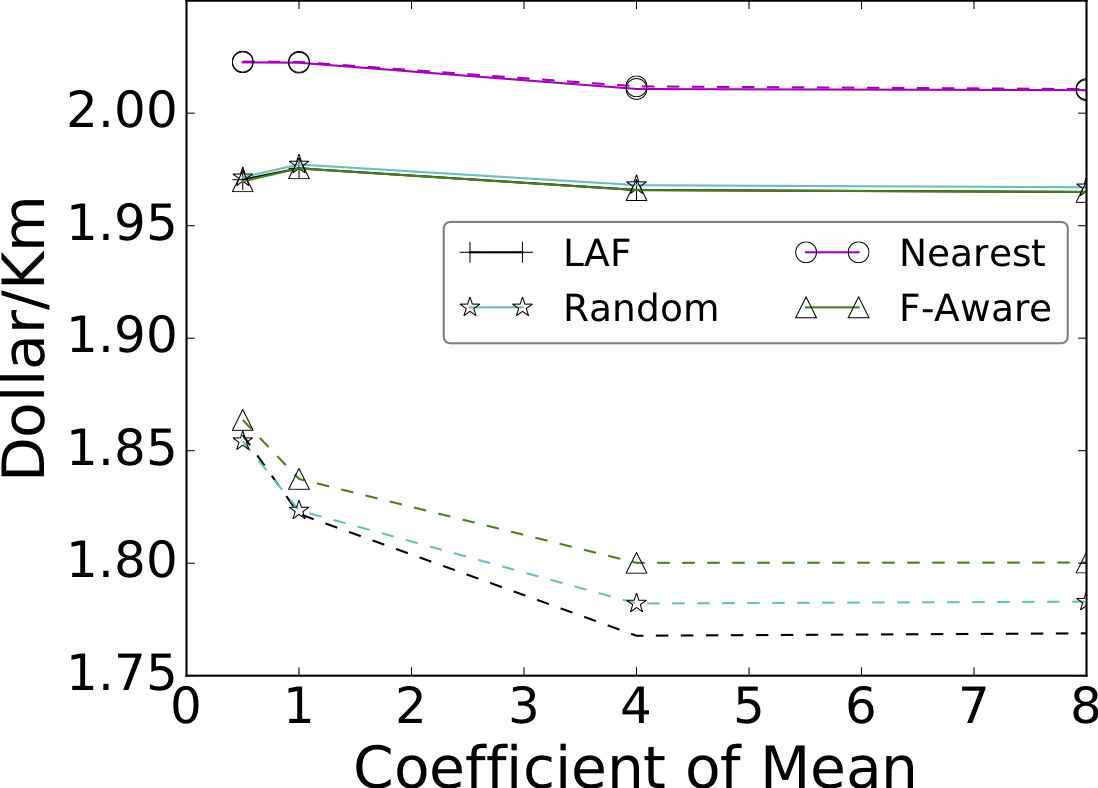}
 			\caption{Earnings of Workers Presented as Dollar per Kilometer}
			\label{fig:discussions_revenue}
		\end{subfigure}
		\caption{Discussions}
	\end{minipage}
\end{figure*}

\subsection{Online Allocation}
We observe the task allocation ratio and unfairness as a function of the window
size and as a function of capacity. Figure~\ref{fig:online} shows our online
allocation related results. For
Figures~\ref{fig:online_tcr}~and~\ref{fig:online_unfairness}, the $x$-axis
represents the window size in terms of minutes. In addition to window size
equals to $0$, the $x$-axis also includes data points starting from $2$
minutes, going up to $64$, doubling at each step. Different series represent
different task allocation algorithms. For
Figures~\ref{fig:online_cap_tcr}~and~\ref{fig:online_cap_unfairness}, the
$x$-axis represents worker capacities. The values are doubled for each data
point, starting from $16$, going up to $512$. For all figures, the $y$-axis
represents a performance metric. In this set, we used $96,000$ tasks.

Figure~\ref{fig:online_tcr} shows the task allocation ratio as a function of
the window size. The red line with cross markers represents offline \emph{F-Aware}
algorithm. All the other series are drawn relative to this line, therefore it
is the $y = 1$ line. We observe that for the instant task allocation, the
difference between different series is negligible. This is because the number
of nominees for each task is very small, and thus the decisions of the
algorithms do not create any difference. On the other hand, when the window
size is increased, we can see that \emph{F-Aware}, and \emph{LAF} performs 
better than the other two approaches and gets closer to the offline
assignment. For the $64$ minutes window, \emph{F-Aware}, and \emph{LAF} allocates $88\%$,
and $85\%$ of the
tasks assignable by offline allocation, whereas \emph{Nearest}, and \emph{Random}
allocation approaches stay at $79\%$ and $76\%$, respectively.

Figure~\ref{fig:online_unfairness} plots unfairness as a function of the
window size using the same setup as Figure~\ref{fig:online_tcr}. 
When smaller windows are used, since tasks are offered to only current nominees,
the unfairness behavior is similar to using smaller $k$ values. For example,
unfairness is $0$, when instant allocation strategy applied, as in unicasting. 
For longer window sizes, 
the results expected to become more similar to the offline setup. For example
The unfairness of offline \emph{F-Aware} is $0.13$, while online \emph{F-Aware}
increases from $0.04$ to $0.11$ when window size increased to $64$ minutes from $2$ minutes.
The most important observation is \emph{F-Aware} performs best among all online algorithms.
For the window size of $64$ minutes online \emph{F-Aware} has $0.11$ unfairness while \emph{LAF},
\emph{Nearest}, and \emph{Random} have $0.15$, $0.32$, $0.33$ respectively. 
We also observe the other three algorithms are even less fair than offline \emph{F-Aware},
for windows larger than $16$ minutes.

Figures~\ref{fig:online_cap_tcr}~and~\ref{fig:online_cap_unfairness} plot the
task allocation ratio and unfairness as a function of worker capacity.
Different series represent different window sizes, including offline and
instant assignments. For all series, \emph{F-Aware} algorithm is used. We make
two main observations. First, the task allocation ratio for smaller window
sizes is higher compared to larger window sizes. The reason
behind this is that, some of the availabilities satisfying tasks expire before
making a decision. At first sight, one might think larger window sizes should
produce closer results to the offline scenario, but this is not the case. In
the offline scenario all the information is known in advance, and decisions
are made before expiration. In contrast, with large windows the availabilities
might expire before processing. Second, larger windows result
in less fair allocation. As we mentioned earlier, smaller windows behave like
batches with smaller $k$ values. As batch-size experiments showed, when a task
offered to less number of workers, fairness increases since sum of total accepted
offers for each worker decreases. 

\subsection{Sensitivity Experiments}
We observe the task allocation ratio and unfairness as a function of the
coefficient of mean. Different series represent different $\Delta T$ values.
Recall that these are the two parameters involved when we used the taxi trips as the tasks, 
and the check-ins as spatio-temporal availabilities of workers. 
The radius added to
a check-in is sampled from a Normal distribution with mean and standard
deviation set to that of the taxi trips. 
For each data point on the $x$-axis,
the mean of this distribution is multiplied with the respective number. For
all experiments, $y$-axis represents a performance metric.

Figure~\ref{fig:sensitivity_tcr} plots the task allocation ratio as a function
of the coefficient of radius. We observe that $\Delta T$ has a great effect on
the task allocation ratio. When coefficient of radius is $1$ and $\Delta T =
1$, $57\%$ of the tasks are allocated, while this number is $82\%$ when
$\Delta T = 8$ for the same coefficient. On the other hand, we cannot observe
the same effect for larger radius values. Consider the $\Delta T = 4$ line.
The task allocation ratio increases by only $0.1\%$ when the coefficient is
increased from $1$ to $4$.

Figure~\ref{fig:sensitivity_unfairness} plots unfairness using the same setup
from Figure~\ref{fig:sensitivity_tcr}. We observe that the difference between
unfairness values is negligible. The peaks are a results of randomness present
in task acceptance. Therefore, we conclude that unfairness is not effected
from the adaption of real-world data to our problem setup.

\subsection{Discussions}
We present two additional experiments that provide insights of our 2-phase
assignment model and the \emph{F-Aware} algorithm.
Figure~\ref{fig:discussion_dist} shows the distribution of workers as a
function of the \fb{\emph{average earning per acceptance}.}
We observe that $800$ workers have
average values between $2\$$ and $4\$$ when \emph{F-Aware} algorithm is used. Although
the most dense areas similar for other approaches as well, we can see that the
standard deviation of them is much higher, which is an indicator of
unfairness.

Unlike traditional crowdsourcing, in spatial crowdsourcing workers have
to physically travel to the source and destination of the task. Therefore, 
dollar earned per traveled kilometer is a good indicator of the what is 
the reward of a worker in return of her labor. In this point we would like
to remind that, batched progressive offer strategy multicasts the offer to
the workers who are most likely to accept it, implicitly workers who are
closer to the task.  

Figure~\ref{fig:discussions_revenue} presents dollar earned per kilometer
as a function of  the coefficient of mean values as in the sensitivity experiments. 
Solid lines represent different assignment algorithms when the batched progressive offer strategy is used,
while dashed lines show the same but when the offers are broadcasted.
Recall that the nearest worker assignment is beneficial
for capturing the spatial aspect of the assignment problem. In case of
broadcasting, represented with dashed lines, we observe the \emph{Nearest} better than other approaches.
Using this approach, workers could make more than $20$ cents per kilometer more compared
to other approaches. On the other hand, we observe the other three approaches benefit
significantly from multicasting. The difference between \emph{Nearest} and \emph{F-Aware}
decreases from $22$ cents to $5$ cents when the system prefers to multicast the offers 
instead of broadcast. Therefore, we could state that \emph{F-Aware} approach captures the spatial
aspect of the problem, with the help of multicasting, as well as allocating each task to \emph{Nearest} worker.

In summary, our experimental evaluation shows that: 
\begin{itemize}
\item \emph{F-Aware} is efficient: It runs around $10^7$ times faster than the \emph{MCF} algorithm and allocates $98.1\%$ of the assignable tasks.
\item \emph{F-Aware} is fair: Unfairness metric of the \emph{LAF} is $2.5\times$ that of \emph{F-Aware} and, it maintains its fairness with increasing capacity, and increasing task count.
\item  Batched progressive offering is useful to limit the assignment ratio,
while not sacrificing the wait time. 
Moreover, it helps to capture the spatial aspect of the problem
by helping to provide competitive dollar per kilometer ratio with \emph{Nearest} worker approach.
\end{itemize}

  \section{Related Work}\label{sec:related}
\noindent We discuss the related work in the areas of crowdsourcing, including spatial
crowdsourcing and fairness.
\smallskip
\\
\textbf{Crowdsourcing.} A wide spectrum of crowdsourcing applications are surveyed in~\cite{ref:csSurvey}.
Crowdsourcing is proposed also for technical tasks such as relational query processing~\cite{ref:crowdDB, ref:deco, ref:qurk}. 
In~\cite{ref:vldb2017}, three crucial 
aspects of crowdsourcing, namely task design, marketplace dynamics, and worker behavior 
are analyzed. However, none of these methods from the literature have considered fairness
among workers.
To the best of our knowledge, this is the first work that applies the
concept and findings of fairness from social psychology
research to crowdsourcing applications.
\smallskip
\\
\textbf{Spatial Crowdsourcing.} Spatial crowdsourcing requires workers to physically travel to
task locations. Earlier work~\cite{ref:realWorld, ref:kazemi, ref:spatial}
extended crowdsourcing to the physical world, with a variety of applications such as 
answering queries~\cite{ref:crowdDB, ref:deco} and serving micro-tasks (such as taking a photo of a monument)~\cite{ref:rdb-sc,
ref:microTask, ref:gMission}. Numerous work has addressed
the maximum task assignment problem~\cite{ref:kazemi,
ref:spatial, ref:kazemiPlus} and its extensions that integrate the reliability of workers
\cite{ref:geoTrust, ref:hassanReliability}. In
\cite{ref:rdb-sc}, maximizing the reliability and spatial diversity are considered
together. Although these works study task allocation in spatial crowdsourcing,
they mostly focus on offline scenarios. In~\cite{ref:taskOnline},
online allocation is performed when only the workers are dynamic. 
In~\cite{ref:microTask}, micro tasks are allocated when both tasks and workers
can appear anywhere, anytime. Different from our work, \cite{ref:hassanMulti} learns 
the workers' acceptance probability in dynamic tasks static workers setup.
Recall, we assume acceptance depends on the hardness level of the tasks
and independent from previous acceptances. Unlike these works, in crowdsourced delivery,
redundant task allocation is not possible.
\rev{Similar to our problem, in the context of crowdsensing, there is a trade off among quality of information (QoI),
budget and time constraints, which requires multi-objective aware task allocation algorithms as well. A recent
work solves this problem with a particle swarm optimization technique that maximizes the aggregated QoI/budget
ratio. A delegation mechanism is used, in case the workers cannot finish 
their allocated task, they may recommend a set of workers from their social network to complete it~\cite{ref:rebecca}. 
This technique serves the same purpose with our 2-phase allocation model.}
Fairness is not in the scope of any of the aforementioned work.
\smallskip
\\
\noindent \textbf{Fairness in Crowdsourcing}. In various application domains,
such as networking~\cite{ref:bandwidth}, staff
scheduling~\cite{ref:staffScheduling}, and resource
allocation~\cite{ref:resource}, fairness is considered to be one of the most
important constraints. In~\cite{ref:servedFairness}, fairness among
customers, \fb{but not workers}, is considered. \fb{Its goal is making sure that the system is
fair to the customer when it is not possible to serve all tasks.} 
\fb{In~\cite{ref:yahia} unfairness is defined as discrimination against individuals, while
in~\cite{ref:papa}, the authors differentiate between various perspectives on fairness and ethics in crowdsourcing. 
They consider distributional fairness as a subjective measure and discuss ethical implications.}
In~\cite{ref:omega}, in addition to the cost
minimization objective, fair allocation of tasks to heterogeneous workers
(workers with different capacities and costs to execute the task) is studied.
The central idea of fairness in~\cite{ref:omega} is to maximize the minimum utility
of all bidders (workers). Utility is defined as the number of allocations. They
also assume that the set of all tasks is known in advance and workers are
predefined. Unlike previous works in spatial crowdsourcing,
\cite{ref:fairAllocation} stresses both the cost incurred by the movement of
the worker and the fairness of the assignment among the workers. A sequence of
sets of spatial tasks are assigned to crowdsourced workers as they arrive. The
one-to-one assignment of tasks to workers is done in mini-batches. 
In their setup, workers are not assigned to multiple tasks. On the 
other hand, to provide redundant task allocation they copy the task,
and allocate one worker for each copy. Besides utilizing
a redundant task allocation strategy, the fairness definition of this work is
different from for our scenario. Yet, in the experimental evaluation we
included the least allocated worker first, \emph{LAF}, which is inspired from this work.
We showed that, according to the state-of-the-art fairness
definition, \emph{LAF} is not fair unless all workers are homogeneous.

  \section{Conclusion}\label{sec:conclusion}
\noindent In this paper, we created a strategy on allocation of \emph{delivery tasks}. 
In this strategy, we use a combined objective 
of maximum task allocation and fair distribution of tasks to workers. 
In our 2-phase allocation model, for each task a set of
nominees are identified using availabilities of workers. The task is offered
to nominees using our \emph{batched-progressive offer strategy}. Once the
candidates for each task are identified, we showed that the problem can be
reduced to Minimum Cost Flow(MCF) problem if fairness is not considered. To
cope with drawbacks of MCF-based solutions in terms of running time and lack of
fairness handling, we introduce our \emph{F-Aware} algorithm. We then
adapt our model to online task allocation and mini-batch task allocation
scenarios. Our evaluation showed that \emph{F-Aware} runs around $10^7 \times$ faster
than the TAR-optimal solution and assigns $96.9\%$ of the tasks that can be
assigned by it. Moreover, F-Aware assigns $18\%$ more tasks than random assignment
approach and is $2.5\times$ more fair than the least allocated worker first assignment strategy.

Our experimental evaluation showed that the distributed fairness criteria can be satisfied with no significant changes in task allocation ratios. The proposed approach of fair allocation of tasks can lead to more sustainable crowdsourced delivery platforms. 
Human perspectives of fairness with quantitative and qualitative surveys, and long-term effects of fair allocation strategies 
in real crowdsourced delivery platforms are among potential future work in this area.
  \bibliographystyle{abbrv} 
  \bibliography{paper}
  \vspace{-1.48cm}
\begin{IEEEbiography}[{\vspace{-0.55cm}\includegraphics[height=1in,clip,keepaspectratio]{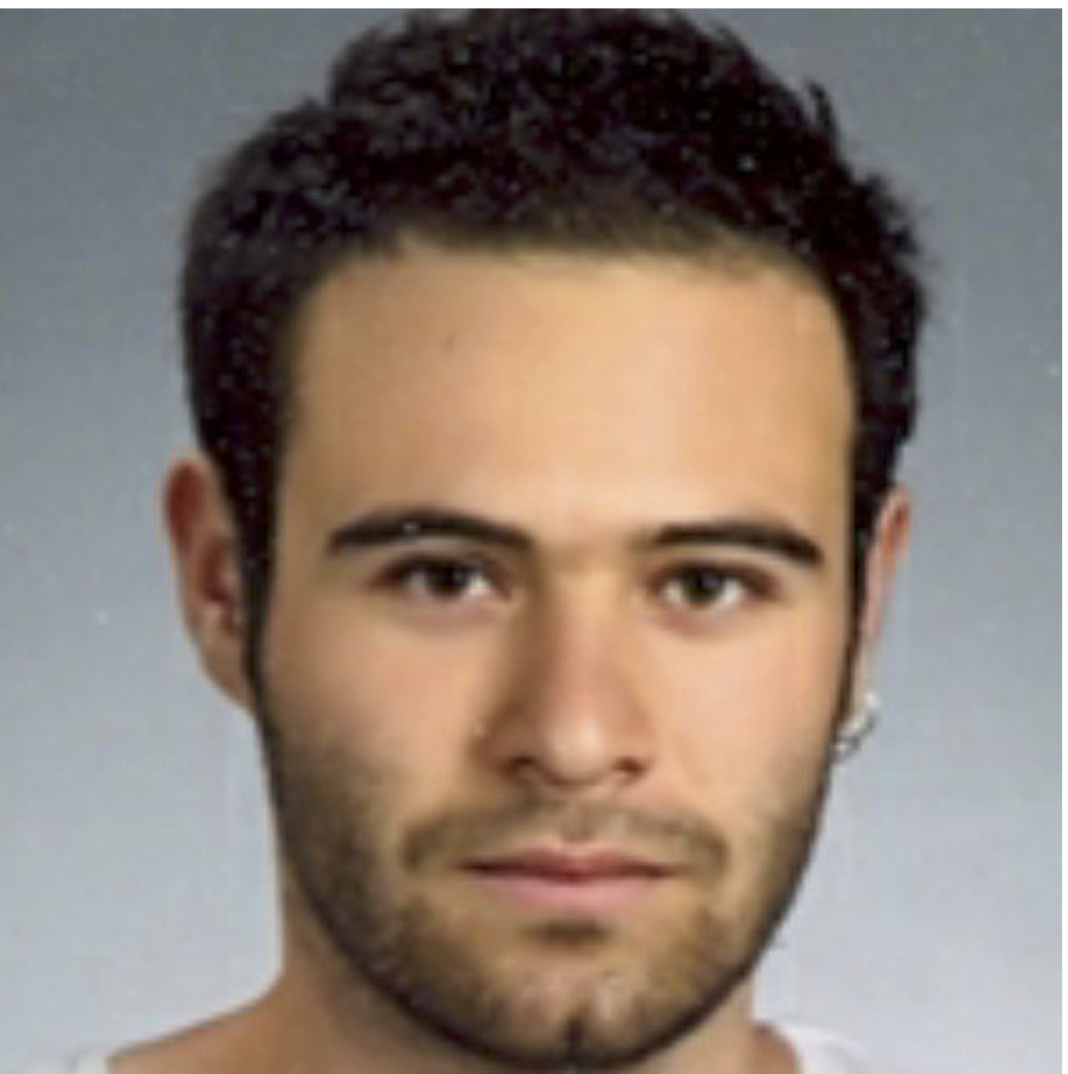}}]
{Fuat Bas{\i}k} is a graduate student in the Department of Computer
Engineering, Bilkent University, Turkey. He holds a M.Sc. degree in Computer Science from Bilkent University. His research interests are in scalable data integration.
\end{IEEEbiography}        
\vspace{-2.00cm}
\begin{IEEEbiography}[{\vspace{-1.00cm}\includegraphics[height=1in,clip,keepaspectratio]{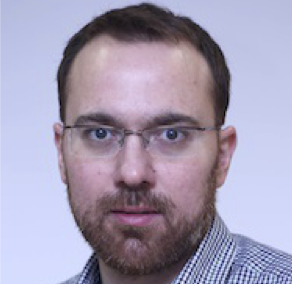}}]
{Bu\u{g}ra Gedik} is an Associate Professor in the Department of Computer Engineering,
Bilkent University, Turkey. He holds a Ph.D. degree in Computer Science from
Georgia Institute of Technology. His research interests are in data-intensive
distributed systems.
\end{IEEEbiography}             
\vspace{-2.25cm}
\begin{IEEEbiography}[{\vspace{-0.55cm}\includegraphics[height=1in,clip,keepaspectratio]{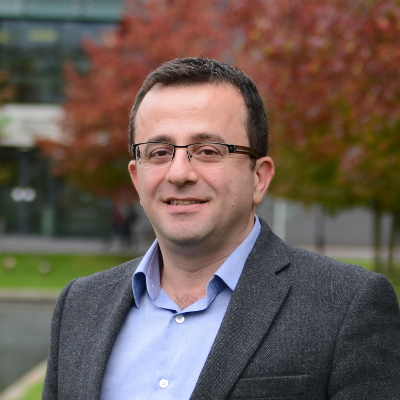}}]
{Hakan Ferhatosmano\u{g}lu} is a Professor in the Department of Science at the University of Warwick. His research is on scalable data management and analytics for multi-dimensional data. He holds a Ph.D. degree in Computer Science from University of California, Santa Barbara. He received research career awards from the US Department of Energy, US National Science Foundation, The Science Academy of Turkey, and Alexander von Humboldt Foundation.
\end{IEEEbiography}
\vspace{-2.00cm}
\begin{IEEEbiography}[{\vspace{-1.00cm}\includegraphics[height=1in,clip,keepaspectratio]{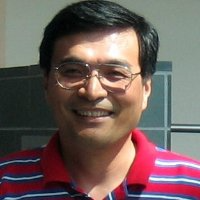}}]
{Kun-Lung Wu} is a Manager at the IBM T. J. Watson
Research Center. His research interests are in big
data systems and applications.
\end{IEEEbiography}

\end{document}